%% file: main.tex
\documentclass[sigconf]{acmart}

\AtBeginDocument{%
  }
\usepackage{amsthm, amsmath, mathtools}
\usepackage{upgreek}
\usepackage{amsfonts}
\usepackage{enumitem}
\usepackage[normalem]{ulem}
\usepackage[linesnumbered,vlined,ruled,commentsnumbered]{algorithm2e}
\usepackage{url}
\usepackage{bm}
\usepackage[labelfont=bf,skip=0pt]{caption,subfig}
\usepackage{flushend}
\usepackage{tabularx}
\usepackage[english]{babel}
\usepackage{graphicx}
\graphicspath{{./figs/}}
\usepackage{thmtools}
\usepackage{listings}
\usepackage{cleveref}
\renewcommand{\paragraph}[1]{\smallskip \noindent {\bf #1}}
\newcommand{\I}{\mathbf{I}}
\newcommand{\dom}{\textsf{dom}}
\newcommand{\Q}{\mathcal{Q}}
\newcommand{\D}{\mathbb{D}}
\newcommand{\Z}{\mathbb{Z}}
\newcommand{\N}{\mathbb{N}}
\renewcommand{\H}{\mathcal{H}}
\newcommand{\ME}{\xi}
\renewcommand{\AE}{\gamma}
\newcommand{\F}{\mathbb{F}}

\newcommand{\eat}[1]{}

\newcommand{\OUT}{\mathsf{OUT}}
\newcommand{\countsize}{\mathrm{count}}
\newcommand{\x}{\mathbf{x}}
\newcommand{\y}{\mathbf{y}}
\newcommand{\s}{\mathbf{s}}
\newcommand{\tdeg}{\widetilde{\mathrm{deg}}}
\renewcommand{\deg}{\mathrm{deg}}
\newcommand{\degg}{\mathrm{mdeg}}
\newcommand{\eps}{\epsilon}
\newcommand{\tOmega}{\widetilde{\Omega}}
\newcommand{\tO}{\widetilde{O}}
\newcommand{\fup}{f^\mathrm{upper}}
\newcommand{\flo}{f^\mathrm{lower}}

\newcommand{\tDelta}{\widetilde{\Delta}}
\newcommand{\PMW}{\textsc{PMW}}
\newcommand{\poly}{\mathop{\mathrm{poly}}}
\newcommand{\RS}{\mathrm{RS}}
\newcommand{\GS}{\mathrm{GS}}
\newcommand{\LS}{\mathrm{LS}}
\newcommand{\cC}{\mathcal{C}}
\newcommand{\bbI}{\mathbb{I}}
\newcommand{\cA}{\mathcal{A}}
\newcommand{\ind}{\mathbf{1}}
\newcommand{\tq}{\widetilde{q}}
\newcommand{\cD}{\mathcal{D}}

\newcommand{\Lap}{\textsf{Lap}}
\newcommand{\TLap}{\textsf{TLap}}
\newcommand{\hn}{\widehat{n}}
\newcommand{\atom}{\textsf{atom}}

\usepackage[colorinlistoftodos,prependcaption,textsize=scriptsize]{todonotes}
\providecommand{\tdotoggle}{1}
\ifnum\tdotoggle=1
\fi

\newif\ifbeforepasinchange

\beforepasinchangefalse

\newcommand{\mytodo}[1]{\ifnum\tdotoggle=1{#1}\fi}


\copyrightyear{2023}
\acmYear{2023}
\setcopyright{acmlicensed}\acmConference[PODS '23]{Proceedings of the 42nd
ACM SIGMOD-SIGACT-SIGAI Symposium on Principles of Database
Systems}{June 18--23, 2023}{Seattle, WA, USA}
\acmBooktitle{Proceedings of the 42nd ACM SIGMOD-SIGACT-SIGAI
Symposium on Principles of Database Systems (PODS '23), June 18--23, 2023,
Seattle, WA, USA}
\acmPrice{15.00}
\acmDOI{10.1145/3584372.3588665}
\acmISBN{979-8-4007-0127-6/23/06}

\begin{document}

\title{Differentially Private Data Release over Multiple Tables}

\author{Badih Ghazi}
\affiliation{%
  \institution{Google Research}
  \country{}
}
\email{badihghazi@gmail.com}

\author{Xiao Hu}
\affiliation{%
  \institution{University of Waterloo}
  \country{}
}
\email{xiaohu@uwaterloo.ca}
\authornote{This work was done while the author was visiting Google Research.}

\author{Ravi Kumar}
\affiliation{%
  \institution{Google Research}
  \country{}
}
\email{ravi.k53@gmail.com}

\author{Pasin Manurangsi}
\affiliation{%
  \institution{Google Research}
  \country{}
}
\email{pasin@google.com}

\begin{abstract}
We study synthetic data release for answering multiple linear queries over a set of database tables in a differentially private way. Two special cases have been considered in the literature: how to release a synthetic dataset for answering multiple linear queries over a single table, and how to release the answer for a single counting (join size) query over a set of database tables. Compared to the single-table case, the join operator makes query answering challenging, since the sensitivity (i.e.,  by how much an individual data record can affect the answer) could be heavily amplified by complex join relationships. 
 
We present an algorithm for the general problem, and prove a lower bound illustrating that our general algorithm achieves parameterized optimality (up to logarithmic factors) on some simple queries (e.g., two-table join queries) in the most commonly-used privacy parameter regimes. For the case of  hierarchical joins, we present a data partition procedure that exploits the concept of {\em uniformized sensitivities} to further improve the utility.
\end{abstract}


\maketitle

\input{intro}

\input{prelim}

\section{Join-As-One Algorithm}
\label{sec:join-as-one}
In this section, we present our {\em join-as-one} algorithm for general join queries, which computes the join results as a single function and then invokes the single-table {\em private multiplicative weights} (PMW) algorithm~\cite{hardt2012simple} (see \Cref{alg:pmw}).  While this is apparently simple, there are many challenging issues in putting everything together.  (All missing proofs are in Appendix~\ref{appendix:join-as-one}.)

\input{two-table}
\input{two-table-lb}
\input{mutil-table}
\input{uniformize}
\input{conclusion}
\bibliographystyle{ACM-Reference-Format}
\bibliography{reference}
\appendix
\input{appendix}

\end{document}

%% file: intro.tex
\section{Introduction}
\label{sec:intro}

 Synthetic data release is a very useful objective in private data analysis. Differential privacy (DP) \cite{dwork2006calibrating,dwork2006our} has emerged as a compelling model that enables us to formally study the tradeoff between utility of released information and the privacy of individuals. In the literature, there has been a large body of work that studies synthetic data release over a single table, e.g., the private multiplicative weight algorithm~\cite{hardt2012simple}, the histogram-based algorithm~\cite{vadhan2017complexity}, the matrix mechanism~\cite{li2010optimizing}, the Bayesian network algorithm \cite{zhang2017privbayes}, and other works on geometric range queries~\cite{li2011efficient, li2012adaptive, dwork2010differential, bun2015differentially, chan2011private, dwork2015pure, cormode2012differentially, huang2021approximate} and datacubes~\cite{ding2011differentially}. 

Data analysis over multiple private tables connected via join operators has been the subject of significant interest within the area of modern database systems. In particular, the challenging question of releasing the join size over a set of private tables has been studied in several recent works including the sensitivity-based framework~\cite{johnson2018towards, dong2021residual,dong2021nearly}, the truncation-based mechanism~\cite{kotsogiannis2019privatesql, dong2022r2t, tao2020computing}, as well as in works on one-to-one joins~\cite{mcsherry2009privacy,proserpio2014calibrating}, and on graph databases~\cite{blocki2013differentially, chen2013recursive}. In practice, multiple queries (as opposed to a single one) are typically issued for data analysis, for example, a large class of linear queries on top of join results with different weights on input tuples, as a generalization of the counting join size query. One might consider answering each query independently but the utility would be very low due to the limited privacy budget, implied by DP composition rules~\cite{dwork2006calibrating, dwork2006our}. Hence the question that we tackle in this paper is: how can we release synthetic data for accurately answering a large class of linear queries over multiple tables privately?

\subsection{Problem Definition}
\noindent{\bf Multi-way Join and Linear Queries.} A (natural) join query can be represented as a hypergraph $\mathcal{H} = (\x, \{\x_1, \dots, \x_m\})$~\cite{abiteboul1995foundations}, where $\x$ is the set of vertices or attributes and each $\x_i \subseteq \x$ is a hyperedge or relation. Let $\dom(x)$ be the domain of attribute $x$. Define $\D = \dom(\x) = \prod_{x \in \x} \dom(x)$, and $\D_i = \dom(\x_i) = \prod_{x \in \x_i} \dom(x)$ for any $i \in [m]$. For attribute(s) $\y$, we use $\pi_\y t$ to denote the value(s) that tuple $t$ displays on attribute(s) $\y$.

Given an instance $\I = \left(R^\I_1, \dots, R^\I_m\right)$ of $\mathcal{H}$, each table $R^\I_i: \D_i \to \mathbb{Z}^{\geq 0}$ 
 is defined as a function from domain $\D_i$ to a non-negative integer, indicating the frequency of each tuple in $\D_i$. This is more general than the setting where each tuple appears at most once in each relation, since it can capture annotated relations~\cite{green2007provenance} with non-negative annotations. The \emph{input size} of $\I$ is defined as the sum of frequencies of all data records, i.e., $n = \sum_{i \in [m]} \sum_{t \in \D_i} R^\I_i(t)$.  

 We encode the natural join as a function $\rho: \times_{i \in [m]}\D_i \to \{0,1\}$, such that for each combination $\vec{t} = (t_1, \dots, t_m) \in \times_{i \in [m]}\D_i$  of tuples,  $\rho(\vec{t}) = 1$ if and only if $\pi_{\x_i \cap \x_j} t_i = \pi_{\x_i \cap \x_j} t_j$ for each pair of $i,j \in [m]$, i.e., $t_i, t_j$ share the same values on the common attributes between $\x_i$ and $\x_j$.
 Given a join query $\mathcal{H}$ and an instance $\I$, the join result is also represented as a function $\textsf{Join}^{\I}: \mathbb{D} \to \mathbb{Z}^{\ge 0}$, such that for any combination $\vec{t} = (t_1,\dots, t_m) \in \times_{i \in [m]} \mathbb{D}_i$,  
 \[\textsf{Join}^{\I} \left(\vec{t}\right) = \rho(\vec{t}) \prod_{i \in [m]} R^\I_i\left(t_i\right).\] 
 The {\em join size} of instance $\I$ over join query $\mathcal{H}$ is therefore defined as
 \[\countsize(\I) = \sum_{\vec{t} \in \times_{i \in [m]}\D_i} \textsf{Join}^\I \left(\vec{t}\right).\]
 
 For each $i \in [m]$, we have a family $\Q_i$ of linear queries defined over $\D_i$ such that for $q \in \Q_i$, $q: \D_i \to [-1,+1]$. Let $\Q  = \times_{i \in [m]}\Q_i$. 
 For each linear query $q = (q_1, \dots, q_m) \in \Q$, the result over instance $\I$ is defined as
 \[ q(\I) = \sum_{\vec{t}= (t_1, \dots, t_m) \in \times_{i \in [m]}\D_i} \rho\left(\vec{t}\right) \prod_{i \in [m]} q_i(t_i) \cdot R^{\I}_i(t_i).\] 
 Our goal is to release a function $\F: \times_{i \in [m]}\D_i \to \mathbb{N}$ such that all queries in $\Q$ can be answered over $\F$ as accurately as possible, i.e., minimizing the $\ell_\infty$-error $\displaystyle{\alpha = \max_{q \in \Q} |q(\I) - q(\F)|}$, where  
 \begin{align*}
    q(\F) &= \sum_{\vec{t} =(t_1, \dots, t_m) \in \times_{i \in [m]}\D_i}  \F(\vec{t}) \prod_{i\in [m]} q_i(t_i). 
 \end{align*}

 We study the data complexity~\cite{vardi1982complexity} of this problem and assume that the size of join query is a constant. When the context is clear, we just drop the superscript $\I$ from $R^\I, \textsf{Join}^\I$. 
 
 \paragraph{DP Setting.} Two DP settings have been studied in the relational model, depending on whether foreign key constraints are
 considered or not. The one considering foreign key constraints assumes the existence of a primary private table, and deleting a tuple in the primary private relation will delete all other tuples referencing it; see~\cite{kotsogiannis2019privatesql, tao2020computing, dong2022r2t}. In this work, we adopt the other notion, which does not consider foreign key constrains, but defines instances to be neighboring if one can be converted into the other by adding/removing a single tuple; this is the same as the notion studied in some previous works~\cite{johnson2018towards, kifer2011no, narayan2012djoin, proserpio2014calibrating}.
 \footnote{Our definition corresponds to the \emph{add/remove} variant of DP where a record is added/removed from a single database $R_i$. Another possible definition is based on \emph{substitution} DP. Add/remove DP implies substitution DP with only a factor of two increase in the privacy parameters; therefore, all of our results also apply to substitution DP.}
 
 \begin{definition}[Neighboring Instances]\label{def:neighboring-addremove}
     A pair $\I = (R_1, \dots, R_m)$ and $\I' = (R'_1, \dots, R'_m)$ of instances are {\em neighboring} if there exists some $i \in [m]$ and $t^* \in \D_i$ such that:
    \begin{itemize}[leftmargin=*]
        \item for any $j \in [m] \smallsetminus \{i\}$, $R_j(t) = R'_j(t)$ for every $t \in \D_j$;
        \item $R_i(t) = R'_i(t)$ for every $t \in \D_i \smallsetminus \{t^*\}$ and $|R_i(t^*) - R'_i(t^*)| = 1$.
    \end{itemize}
 \end{definition}


\begin{definition}[Differential Privacy~\cite{dwork2006calibrating,dwork2006our}]
    For $\epsilon,\delta>0$, an algorithm $\mathcal{A}$ is \emph{$(\epsilon, \delta)$-differentially private} (denoted by $(\epsilon, \delta)$-DP) if for any pair $\I$, $\I'$ of neighboring instances and any subset $\mathcal{S}$ of outputs,
   $\Pr(\mathcal{A}(\I) \in \mathcal{S}) \le e^\epsilon \cdot \Pr(\mathcal{A}(\I') \in \mathcal{S}) + \delta$.
\end{definition}

\paragraph{Notation.}
For simplicity of presentation, we henceforth assume throughout that $0 < \eps \leq O(1)$ and $0 \leq \delta \leq 1/2$. Furthermore, all lower bounds below hold against an algorithm that achieves the stated error $\alpha$ \emph{with probability $1 - \beta$ for some sufficiently small constant $\beta > 0$}; we will omit mentioning this probabilistic part for brevity.
For notational ease, we define the following quantities:
\begin{align*}
    &\flo(\mathbb{D}, \Q, \epsilon) = \sqrt{\frac{1}{\epsilon} \cdot
\sqrt{\log |\D|}}, \mbox{ and } \\
    &\fup(\mathbb{D}, \Q, \epsilon, \delta) = \flo(\mathbb{D}, \Q, \epsilon) \cdot \sqrt{\log |\Q| \cdot \log 1/\delta}.
\end{align*}
When $\D, \Q, \eps, \delta$ are clear from context, we will omit them.  Let $\lambda = \frac{1}{\epsilon} \log \frac{1}{\delta}$, which is a commonly-used parameter in this paper.

\subsection{Prior Work}
We first review the problem of releasing synthetic data for a single table, and mention two important results in the literature. In the single table case, nearly tight upper and lower bounds are known:

\begin{theorem}[\cite{hardt2012simple}]
\label{the:single-table}
    For a single table $R$ of at most $n$ records, a family $\Q$ of linear queries, and $\epsilon, \delta > 0$, there exists an algorithm that is $(\epsilon, \delta)$-DP, and with probability at least $1- 1/\mathrm{poly}(|\Q|)$ produces $\F$ such that all queries in $\Q$ can be answered to within error $\displaystyle{\alpha =O\left(\sqrt{n} \cdot \fup\right)}$ using $\F$.
\end{theorem}

\ifbeforepasinchange
\else


\begin{theorem}[\cite{bun2018fingerprinting}]
\label{the:lb-single-table}
    For every sufficiently small $\eps > 0$, sufficiently large $n$, $n_D \geq n^{O(1)}$ and $n_Q \geq (n \cdot \log n_D)^{O(1)}$, there exists a family $\Q$ of queries of size $n_Q$ on a domain $\D$ of size $n_D$ such that any $(\epsilon, 1/n^{\omega(1)})$-DP algorithm that takes as input a database of size at most $n$, and outputs an approximate answer to each query in $\Q$ to within error $\alpha$ must satisfy $\displaystyle{\alpha \geq \tOmega\left(\min\left\{n, \sqrt{n} \cdot \flo \right\}\right)}$.
\end{theorem}

\fi

Another related problem that has been widely studied by the database community is how to release the join size of a query privately, i.e., $\countsize(\I)$ for every input instance $\I$.
Note that $\countsize(\I)$ is essentially a special linear query $q = (q_1, \dots, q_m)$ with $q_i: \D_i \to \{+1\}$ for every $i \in [m]$. A popular approach of answering the counting join size query is based on
the sensitivity framework, which first computes $\countsize(\I)$, then computes the sensitivity of the query (measuring the difference between the query answers on
neighboring database instances), and releases a noise-masked query answer, where the noise is
drawn from some zero-mean 
distribution calibrated appropriately according to the sensitivity. It is known that the local sensitivity of counting join size query, defined as 
\[\LS_\countsize(\I) = \max_{\I': (\I,\I') \textrm{ are neighboring }} |\countsize(\I) - \countsize(\I')|,\] 
where the maximum is over all neighboring instances $\I'$ of $\I$, cannot be directly used for calibrating noise, as $\LS_\countsize(\I), \LS_\countsize(\I')$ for a pair  $\I,\I'$ of neighboring databases can be used to distinguish them.  Global sensitivity (e.g.,~\cite{dwork2006calibrating}) defined as 
\[\GS_\countsize(\I) = \max_{\I} \LS_\countsize(\I),\]
where the maximum is over all instances $\I$, could be used but the utility is unsatisfactory in many instances. Smooth upper bounds on local sensitivity~\cite{nissim2007smooth} have instead been considered; they offer 
 much better utility than global sensitivity.  Examples include {\em smooth sensitivity}, which is the smallest smooth upper bound but usually cannot be efficiently computed, and {\em residual sensitivity}~\cite{dong2021residual}, which is a constant-approximation of smooth sensitivity but can be efficiently computed and used in practice. See Section~\ref{sec:join-as-one} for formal details. 

We note that the sensitivity-based framework can be adapted to answering any single linear query, as long as the sensitivity of the specific linear query is correctly computed. However, we are not interested in answering each linear query individually, since the privacy budget (implied by DP composition) would blow up with the number of queries to be answered, which would be too costly when the query space is extremely large. Instead, we aim to release a synthetic dataset such that any arbitrary linear query can be freely answered over it while preserving DP. In building our data release algorithm, we also resort to the existing sensitivities derived for the counting join size query but in a more careful way. 

In the remaining of this paper, when we mention ``sensitivity'' without a specific function, it should be assumed that the function is counting join size of an input instance, i.e., $\countsize(\I)$.

%
%

\subsection{Our Results}
%
%
We first present the basic {\em join-as-one} approach for releasing a synthetic dataset for multi-table queries, which  computes the join results as a single table and uses existing algorithms for releasing synthetic dataset for a single table.  The error will be a function of the join size and the residual sensitivity of $\countsize(\cdot)$. 

\ifbeforepasinchange
\else


\begin{theorem}
\label{the:ub-multi-table}
     For any join query $\mathcal{H}$, an instance $\I$, a family $\Q$ of linear queries, and $\epsilon > 0$, $\delta > 0$, there exists an $(\epsilon, \delta)$-DP algorithm that with probability at least $1- 1/\textsf{poly}(|\Q|)$ produces $\F$ that can be used to answer all queries in $\Q$ within error:
    \[\alpha = O\left(\left(\sqrt{\countsize(\I) \cdot \RS^\beta_{\countsize}(\I)} + \RS^\beta_{\countsize}(\I) \cdot \sqrt{\lambda} \right) \cdot \fup \right), \]
    where $\countsize(\I)$ is the join size of $\mathcal{H}$ over $\I$ and $\RS^\beta_\countsize(\I)$ is the $\beta$-residual sensitivity of $\I$ for $\beta = 1/\lambda$.
\end{theorem}

\fi

We next present the parameterized optimality with respect to the join size and the local sensitivity of $\countsize(\cdot)$.

\ifbeforepasinchange
\else


\begin{theorem}
\label{the:lb-multi-table}
    Given arbitrary parameters $\OUT \geq \Delta > 0$ such that $\OUT / \Delta$ is sufficiently large and a join query $\mathcal{H}$,
    for every sufficiently small $\eps > 0$, $n_D \geq \OUT^{O(1)}$ and $n_Q \geq (\OUT \cdot \log n_D)^{O(1)}$, there exists a family $\Q$ of queries of size $n_Q$ on domain $\D$ of size $n_D$ such that any $(\epsilon, 1/n^{\omega(1)})$-DP  algorithm that takes as input a multi-table database over $\mathcal{H}$ of input size at most $n$, join size $\OUT$ and local sensitivity $\Delta$, and outputs an approximate answer to each query in $\Q$ to within error $\alpha$ must satisfy  \[\alpha \ge \tOmega\left(\min\left\{\OUT, \sqrt{\OUT \cdot \Delta} \cdot \flo\right\}\right).\]
\end{theorem}

\fi

 For the typical setting with $\epsilon = \Omega(1), \delta = O\left(\frac{1}{n^c}\right)$ for some constant $c$, $|\Q| \leq n^{O(1)}$ and $\lambda = O\left(\log n\right)$. 
 
 The gap between the upper bound in Theorem~\ref{the:ub-multi-table} and the lower bound in Theorem~\ref{the:lb-multi-table}---ignoring poly-logarithmic factors---boils down to the gap between smooth sensitivity and local sensitivity used in answering $\countsize(\cdot)$ in a private way, which appears in answering such a single query in a different form.


From the upper bound in terms of residual sensitivity, we exploit the idea of {\em uniformized  sensitivity}\footnote{A similar idea of {\em uniformizing degrees} of join values, i.e., how many tuples a join value appears in a relation, has been used in query processing~\cite{joglekar2015s}, but we use it in a more complicated scenario to achieve uniform sensitivity.} 
to further improve Theorem~\ref{the:ub-multi-table}, by using a more fine-grained partition of input instances. For the class of hierarchical queries, which has been widely studied in the context of various database problems~\cite{dalvi2007efficient, fink2016dichotomies, berkholz2017answering, hu2022temporal}, we obtain a more fine-grained parameterized upper bound in Theorem~\ref{the:up-hierarchical-uniformize} and lower bound in Theorem~\ref{the:lb-hierarhical-uniformize}. In the main text, we include Theorem~\ref{the:up-2table-uniformize} and Theorem~\ref{the:lb-instance-optimality-2table} for the basic two-table join to illustrate the high-level idea.  We defer all details to Section~\ref{sec:uniformization}.

%% file: prelim.tex
\section{Preliminaries}

We introduce some commonly-used concepts, terminologies, and primitives used in this paper.

\paragraph{Notation.} For random variables $X, Y$ and scalars $\eps >0, \delta \in [0, 1)$, we use $X \approx_{(\eps, \delta)} Y$ for short if $\Pr[X \in S] \leq e^\eps \cdot \Pr[Y \in S] + \delta$ and $\Pr[Y \in S] \leq e^\eps \cdot \Pr[X \in S] + \delta$ for all sets $S$ of outcomes. For convenience, we sometimes write the distribution in place of a random variable drawn from that distribution. E.g., $a + \cD$ is a shorthand for $a + X$, where $X$ is a random variable with distribution $\cD$ (denoted $X \sim \cD$).

\paragraph{(Truncated) Laplace Distribution.}
The \emph{Laplace distribution} with parameter $b$, denoted  $\Lap_b$, has probability density function (PDF) $\Lap_b(x) \propto e^{-|x|/b}$. The \emph{shifted and truncated Laplace distribution} with parameters $b$ and $\tau$, denoted $\TLap_b^\tau$, is the distribution supported on $[0, 2\tau]$ whose PDF over the support satisfies $\TLap_{b}^{\tau}(x) \propto e^{-|x - \tau|/b}$. 
The DP guarantees of the Laplace and truncated Laplace distributions are well-known (e.g.,~\cite{dwork2006calibrating, GengDGK20, GhaziKKMZ22}): for any pair  $u,v$ with $|u -v| \le \Delta$, $u + \Lap_{\Delta/\eps} \approx_{(\epsilon, 0)} v + \Lap_{\Delta/\eps}$
and 
$u + \TLap_{\Delta/\eps}^{\tau} \approx_{(\epsilon, \delta)} v + \TLap_{\Delta/\eps}^{\tau}$
for $\tau = \tau(\eps, \delta, \Delta) = \frac{\Delta}{\eps}\ln\left(1 + \frac{e^\eps - 1}{\delta}\right)$. Note here that $\tau(\eps, \delta, \Delta) \leq O(\Delta \cdot \lambda)$ when $\epsilon$ is a constant.



\paragraph{Exponential Mechanism.}
The exponential mechanism (EM)~\cite{McSherryT07} selects a ``good'' candidate from a set $\cC$ of candidates. The goodness is defined by a \emph{scoring function} $s(\I, c)$ for an input dataset $\I$ and a candidate $c \in \cC$ and is assumed to have sensitivity at most one. Then, the EM algorithm consists of sampling each candidate $c \in \cC$ with probability $\propto \exp\left(-0.5\eps \cdot s(\I, c)\right)$; this algorithm is $(\eps, 0)$-DP.
%
%
%


%% file: two-table.tex
\subsection{Algorithms for Two-Table Query}
We start with the simplest two-table query. Assume the join query $\mathcal{H}$ is given by the hypergraph with vertices $\x = \{A,B,C\}$ and hyperedges $\x_1 = \{A,B\}$, $\x_2 = \{B,C\}$. For $i \in \{1,2\}$, we define the \emph{degree} of a join value $b \in \dom(B)$ in $R_i$ (i.e., the frequencies of data records displaying join value $b$ in $R_i$) to be $\displaystyle{\deg_{i,B}(b) = \sum_{t \in \D_i:\pi_B t=b} R_i(t)}$. The \emph{maximum degree} of any join value in $\dom(B)$ is denoted as 
$\displaystyle{\Delta = \max_{b \in \dom(B)} \max\{\deg_{1,B}(b), \deg_{2,B}(b)\}}$; note that $\Delta = \LS_\countsize(\I)$. 

\begin{figure}[t]
   \centering
    \includegraphics[scale=1.0]{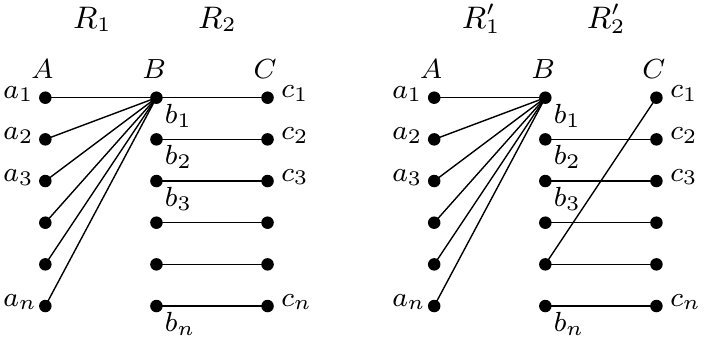}
    \caption{A pair of neighboring instances $\I$ (left) and $\I'$ (right) for two-table join with $\dom(A) = \{a_1, \dots,a_n\}$, $\dom(B) = \{b_1,\dots,b_n\}$, and $\dom(C) = \{c_1, \dots,c_n\}$. }
    \label{fig:counter-example}
\end{figure}

\paragraph{A Natural (but Flawed) Idea.} Consider the following algorithm:
\begin{itemize}[leftmargin=*]
    \item Compute the join result $J = \textrm{Join}^\I$ and release a synthetic dataset $\tilde{J}_1$ for $J$ by invoking the single-table PMW algorithm;
\end{itemize}
We now briefly explain why this algorithm violates DP. Let $\tilde{J}, \tilde{J'}$ be the synthetic dataset released by this algorithm for two neighboring databases $\I,\I'$ respectively. By the property of the single-table PMW algorithm, the total number of tuples over the whole domain stays unchanged, i.e., $\countsize(\I) = \sum_{\vec{t} \in \mathbb{D}_1 \times \mathbb{D}_2} \tilde{J}(\vec{t})$ and $\countsize(\I') = \sum_{\vec{t} \in \mathbb{D}_1 \times \mathbb{D}_2} \tilde{J'}(\vec{t})$. 
However, $\I, \I'$ could have dramatically different join sizes: Figure~\ref{fig:counter-example} shows two neighboring databases with join sizes $n$ and $0$ respectively.  Thus, $\tilde{J}, \tilde{J'}$ can be used by an adversary to distinguish $\I, \I'$, which makes the algorithm not DP.

\paragraph{Another Natural (but Still Flawed) Idea.} To remedy the leakage in the above algorithm, we seek to protect the total number of tuples in the released dataset, say, by padding with dummy tuples:
\begin{enumerate}[leftmargin=*]
    \item Compute the join result $J = \textsf{Join}^\I$ and release a synthetic dataset $\tilde{J}_1$ for $J$ by invoking the single-table PMW algorithm;
    \item $\tDelta \gets \Delta + \TLap_{2/\eps}^{\tau(\eps/2,\delta/2,1)}$; 
    \item $\tilde{J_2} \gets$ random sample of $\eta$ records from $\mathbb{D}_1 \times \mathbb{D}_2$, where
    $\eta \sim \TLap_{2\tDelta/\eps}^{\tau(\eps/2,\delta/2,\tDelta)}$;    
    \item Release the union of these two datasets: $\F = \tilde{J}_1 \cup \tilde{J}_2$;
\end{enumerate}
As noted, the local sensitivity cannot be directly used for privately releasing the join size, however, the global sensitivity of $\LS_\countsize(\cdot)$ is $1$. Hence, the second step is to obtain an upper bound on the local sensitivity, which will be used to calibrate the noise needed to mask the join size of $\I$. More specifically, for any neighboring databases $\I, \I'$, let $\mathbb{F}, \mathbb{F}'$ be the synthetic datasets released for $\I,\I'$ respectively. As mentioned, $\sum_{t \in \mathbb{D}_1 \times \mathbb{D}_2} \mathbb{F}(t) \approx_{(\epsilon, \delta)} \sum_{t \in \mathbb{D}_1 \times \mathbb{D}_2} \mathbb{F}'(t)$, which resolves the information leakage of the join size. However, it turns out that this approach can still lead to the distinguishability of neighboring databases $\I,\I'$.  Indeed, consider the following example.
\begin{example}
\label{exp:dp}
Consider neighboring instances $\I,\I'$ in Figure~\ref{fig:counter-example}. Assume $\epsilon = \Theta(1)$ and $\delta = O(1/n^c)$ for some constant $c$. Let $\tilde{J}_1, \tilde{J}'_1$ be the synthetic datasets released for $J,J'$ respectively. As $\tilde{J}(\vec{t})=0$ for each $\vec{t} \in \mathbb{D}_1 \times \mathbb{D}_2$, $\tilde{J'}(\vec{t})=0$ also holds for every $\vec{t} \in \mathbb{D}_1 \times \mathbb{D}_2$. Let $\mathbb{D}' = \left(\dom(A) \times \{b_1\}\right) \times \left(\{b_1,c_1\}\right)$. In contrast, with probability at least $1 -1/\textsf{poly}(|\Q|)$, there exists some constant $c_1$ such that
$\sum_{\vec{t} \in \mathbb{D}'} \tilde{J}_1 \left(\vec{t}\right) \ge n - c_1 \cdot \sqrt{n} \cdot \fup$, for answering $\countsize(\cdot)$ within an error of $O(\sqrt{n} \cdot \fup)$. Moreover, $\Delta = n$, hence $\Tilde{\Delta} = O(n \log n)$ and $\eta = O(n \log^2 n)$. The probability that no tuple in $\mathbb{D}'$ is sampled by $\tilde{J}'_2$ is
$(1 -n/n^3)^{c_2 \cdot n \log^2 n} = (1 - 1/n^2)^{n^2 \cdot \frac{c_2 \log n^2}{n}} > 1/e$ for some constant $c_2$. Together, $\Pr[\sum_{\vec{t} \in \mathbb{D}'} \mathbb{F}(\vec{t}) = 0] < 1/\poly(|\Q|)$ and $\Pr[\sum_{\vec{t} \in \mathbb{D}'} \mathbb{F}'(\vec{t}) = 0] > 1/e$, 
thus violating the DP condition $1/e < 1/\poly(|\Q|) \cdot e^{\Theta(1)} + O(1/n^c)$.  
\end{example}
\paragraph{Our Algorithm.} The main insight in the idea that works is to change the order of the two steps in the previous flawed version. More specifically, we first pad dummy tuples to join results and then release the synthetic dataset for the ``noisy'' join results. 
\begin{enumerate}[leftmargin=*]
    \item $\tDelta \gets \Delta + \TLap_{2/\eps}^{\tau(\eps/2,\delta/2,1)}$; 
    \item $\tilde{J_2} \gets$ random sample of $\eta$ records from $\mathbb{D}_1 \times \mathbb{D}_2$, where
    $\eta \sim \TLap_{2\tDelta/\eps}^{\tau(\eps/2,\delta/2,\tDelta)}$;    
    \item We compute the join result $J = \textrm{Join}^\I$ and release a synthetic dataset $\F$ for $J \cup \tilde{J_2}$ by invoking the single-table PMW algorithm;
\end{enumerate}
\begin{algorithm}[t]
\caption{{\sc TwoTable}$(\I = \{R_1, R_2\})$}
\label{alg:two-table}
$\tDelta \gets \Delta + \TLap_{2/\eps}^{\tau(\eps/2,\delta/2,1)}$\;
\Return $\textsc{PMW}_{\eps/2,\delta/2,\tDelta}(\I)$; \hfill $\blacktriangleright$ \Cref{alg:pmw}\;
\end{algorithm}
\begin{algorithm}[t]
\caption{{\sc PMW}$_{\eps, \delta, \tDelta}(\I = \{R_1, \dots, R_m\})$}
\label{alg:pmw}
$\widehat{n} = \countsize(\I) + \TLap_{2\tDelta/\eps}^{\tau(\eps/2,\delta/2,\tDelta)}$\; 
$\F_0 \gets \widehat{n}$ times uniform distribution over $\D$: $\F_0(x) = \frac{\widehat{n}}{|\D|}$\;
$\eps' \leftarrow \frac{\eps}{16 \sqrt{k \cdot \log(1/\delta)}}$\;
\ForEach{$i = 1, \dots, k$}{
    Sample a query $q_i \in \Q$ using the $\eps'$-DP EM with score function $s_i(\I,q) = \frac{1}{\tDelta} \cdot |q(\F_{i-1}) - q(\I)|$\;
    $m_i \gets q_i(\I) + \Lap_{\tDelta/\eps'}$\;
    Update for each $x \in \D$: $\F_i(x) \propto \F_{i-1}(x) \times \exp\left(q_i(x) \cdot \left(m_i - q_i(\F_{i-1})\right) \cdot \frac{1}{2 \widehat{n}}\right)$\;
}
\Return $\textsf{Avg}_{i=1}^k \F_i$\;
\end{algorithm}
We observe that this approach can be further simplified by releasing a synthetic dataset for the join result $J$ directly by invoking the single-table PMW algorithm, but starting from an initial uniform distribution parameterized by $\hat{n} = \countsize(\I) + \eta$.  Algorithm~\ref{alg:two-table} contains a complete description for releasing synthetic dataset for two-table joins.  Since $\widehat{n}$ is private and the single-table algorithm~\cite{hardt2012simple} releases a private synthetic dataset for $J$, Algorithm~\ref{alg:two-table} also satisfies DP, as stated in Lemma~\ref{lem:two-table-DP}. 

\begin{lemma}
\label{lem:two-table-DP}
    Algorithm~\ref{alg:two-table} is $(\epsilon, \delta)$-DP.
\end{lemma}

\paragraph{Error Analysis.} As shown in Appendix~\ref{appendix:single-table}, with probability $1 - 1/\textsf{poly}(|\Q|)$, Algorithm~\ref{alg:pmw} returns a synthetic dataset 
$\F$ such that every linear query in $\Q$ can be answered over $\F$ within error (omitting $\fup$):
$O\left(\sqrt{\countsize(\I) \cdot (\Delta + \lambda)} + (\Delta + \lambda)\cdot \sqrt{\lambda} \right)$. Then, we arrive at:

\begin{theorem}
\label{the:up-2table}
     For any two-table instance $\I$, a family $\Q$ of linear queries, and $\epsilon > 0$, $\delta > 0$, there exists an algorithm that is $(\epsilon, \delta)$-DP, and with probability at least $1- 1/\textsf{poly}(|\Q|)$ produces $\F$ such that all linear queries in $\Q$ can be answered to within error:
    \[\alpha = O\left((\sqrt{\countsize(\I) \cdot (\Delta + \lambda)} + (\Delta + \lambda)\cdot \sqrt{\lambda}) \cdot \fup\right),\]
    where $\countsize(\I)$ is the join size of $\I$, and $\Delta = LS_\countsize(\I)$.
\end{theorem}

%% file: two-table-lb.tex
\subsection{Lower Bounds for Two-Table Join}



The first lower bound is based on the local sensitivity:

\begin{theorem}
\label{the:lb-two-table-1}
     For any $\Delta > 0$,
     there exists a family $\Q$ of queries  such that any $(\epsilon, \delta)$-DP algorithm that takes as input a database 
     $\I$ with local sensitivity at most $\Delta$
      and outputs an approximate answer to each query in $\Q$ to within error $\alpha$, must satisfy $\alpha \ge \Omega\left(\Delta\right)$.
\end{theorem}

Our second lower bound is via a reduction to the single-table case: we create a two-table instance where $R_1$ encodes the single-table and $R_2$ ``amplifies'' both the sensitivity and the join size by a factor of $\Delta$. This eventually results in the following lower bound:

\ifbeforepasinchange
\else


\begin{theorem}
\label{the:lb-two-table-2}
    Given arbitrary parameters $\OUT \geq \Delta > 0$ such that $\OUT / \Delta$ is sufficiently large,
    for every sufficiently small $\eps > 0$, $n_D \geq \OUT^{O(1)}$ and $n_Q \geq (\OUT \cdot \log n_D)^{O(1)}$, there exists a family $\Q$ of queries of size $n_Q$ on domain $\D$ of size $n_D$ such that any $(\epsilon, 1/n^{\omega(1)})$-DP  algorithm that takes as input a two-table database over $\mathcal{H}$ of input size at most $n$, join size $\OUT$ and local sensitivity $\Delta$, and outputs an approximate answer to each query in $\Q$ to within error $\alpha$ must satisfy  \[\alpha \ge \tOmega\left(\min\left\{\OUT, \sqrt{\OUT \cdot \Delta} \cdot \flo\right\}\right).\]
\end{theorem}

\fi

\begin{figure}[t]
    \centering
    \includegraphics[scale=0.77]{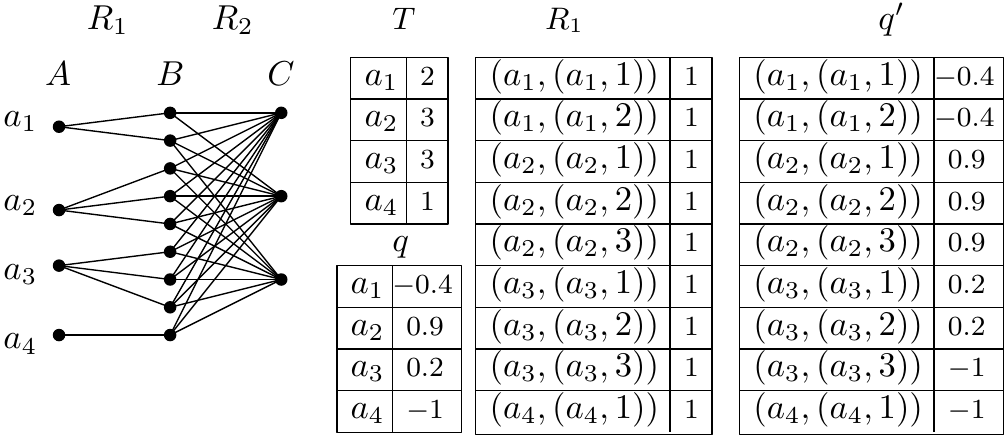}
    \caption{A hard instance constructed for two-table join, with $n=9$, $\Delta=3$ and $\OUT= 27$. }
    \label{fig:lb-two-table}
\end{figure}
\begin{proof}
Let $n= \frac{\OUT}{\Delta}$. From Theorem~\ref{the:lb-single-table}, there exists a set $\Q_\textsf{one}$ of queries on domain $\D$ for which any $(\epsilon, \delta)$-DP algorithm that takes as input a single-table database $T \in \D$ and outputs an approximate answer to each query in $\Q_\textsf{one}$ within error $\alpha$ requires that $\alpha \ge \tilde{\Omega}\left(\min\left\{n, \sqrt{n} \cdot \flo(\mathbb{D}, \Q_\textsf{one}, \epsilon)\right\}\right)$.
For an arbitrary single-table database $T: \D \to \Z^+$, we construct
a two-table instance $\I$ of join size $\OUT$, and local sensitivity $\Delta$ as follows:

\begin{itemize}[leftmargin=*]
    \item Set $\dom(A) = \D$, $\dom(B) = \D \times [n]$, and $\dom(C) = [\Delta]$.
    \item Let $R_1(a, (b_1, b_2)) = \ind[a = b_1 \wedge b_2 \leq T(a)]$ for all $a \in \dom(A)$ and $(b_1, b_2) \in \dom(B)$.
    \item Let $R_2(b, c) = 1$ for all $b \in \dom(B)$ and $c \in \dom(C)$.
\end{itemize}

%
It can be easily checked that $\I$ has join size at most $\OUT$ and local sensitivity $\Delta$, and that two neighboring databases $T, T'$ result in neighboring instances $\I, \I'$.
Finally, let $\Q_1$ contain queries from $\Q_\textsf{one}$ applied on its first attribute (i.e., $\Q_1 := \{q \circ \pi_A \mid q \in \Q_\textsf{one}\}$), and let $\Q_2$ contain only a single query $q_\textsf{all-one}: \D_2 \to \{+1\}$.
An example is illustrated in 
Figure~\ref{fig:lb-two-table}. 

Our lower bound argument is a reduction to the single-table case. 
Let $\mathbb{I}(\OUT, \Delta)$ be the set of all instances of join size $\OUT$ and local sensitivity $\Delta$. Let $\mathcal{A}$ be an algorithm that takes any two-table instance in $\mathbb{I}(\OUT, \Delta)$, and outputs an approximate answer to each query in $\Q$ within error $\alpha'$.  For each query $q \in \Q_\textsf{one}$, let $q' = (q \circ \pi_A, q_\textsf{all-one})$ be its corresponding query in $\Q$. Let $\tq'(\I)$ be the approximate answer for query $\tq'$. We then return $\displaystyle{\tq(T) = \tq'(\I) / \Delta}$ as an approximate answer to $q(T)$.

We first note that this algorithm for answering queries in $\Q_\textsf{one}$ is $(\epsilon, \delta)$-DP due to the post-processing property of DP.  The error guarantee follows immediately from the observation that $q'(\I) = \Delta \cdot q(T)$.
%
%
Therefore, from Theorem~\ref{the:lb-single-table}, we can conclude that 
\begin{align*}
\alpha' &\geq \Delta \cdot \tOmega(\min\{n, \sqrt{n} \cdot \flo(\mathbb{D}, \Q_\textsf{one}, \epsilon)\}) \\ &= \tOmega\left(\min\{\OUT, \sqrt{\OUT} \cdot \sqrt{\Delta} \cdot \flo\}\right). \qedhere
\end{align*}
\end{proof}




%% file: mutil-table.tex
\subsection{Multi-Table Join}

\begin{algorithm}[t]
\caption{{\sc MultiTable}$(\I)$}
\label{alg:multi-table}
$\beta \gets 1/\lambda$\;
$\tDelta \gets \RS^\beta_\countsize(\I) \cdot e^{\TLap_{2\beta/\eps}^{\tau(\eps/2,\delta/2,\beta)}}$\;
\Return $\textsc{PMW}_{\eps/2,\delta/2,\tDelta}(\I)$\; 
\end{algorithm} 

We next extend the join-as-one approach to multi-table queries. First, notice that \Cref{alg:two-table} does not work in the multi-table setting because $\LS_{\countsize}(\cdot)$ may itself have a large global sensitivity (unlike the two-table case where the global sensitivity of $\LS_{\countsize}(\cdot)$ is $1$.) Hence, we will have to find other ways to perturb the join size, and then use this noisy join size to parameterize the initial uniform distribution of the single-table PMW algorithm. As noted, several previous works have proposed various smooth upper bounds on the local sensitivity~\cite{nissim2007smooth}: For $\beta > 0$, $S^\beta(\cdot)$ is a $\beta$-smooth upper bound on local sensitivity, if (1) $S^\beta(\I) \ge \LS_\countsize(\I)$ for every instance $\I$; and (2) for any pair of neighboring instances $\I,\I'$, $S^\beta(\I') \le e^\beta S^\beta(\I)$.

The smallest smooth upper bound on local sensitivity is denoted as {\em smooth sensitivity}. However, as observed by~\cite{dong2021residual}, it takes $n^{O(\log n)}$ time to compute smooth sensitivity for $\countsize(\cdot)$, which is very expensive especially when the input size $n$ of underlying instance is large. Fortunately, we do not necessarily need smooth sensitivity:  any smooth upper bound on local sensitivity can be used for perturbing the join size. In building our data release algorithm, we use {\em residual sensitivity}, which is a constant-approximation of smooth sensitivity, and more importantly can be computed in time polynomial in $n$.

To introduce the definition of residual sensitivity, we need the following terminology. 
Given a join query $\mathcal{H} = (\x, \{\x_1,\dots,\x_m\})$ 
and a subset $E \subseteq [m]$ of relations, 
its \emph{boundary}, denoted as $\partial_{E}$, is the set of attributes that belong to relations both in and out of $E$, i.e.,
$\partial E = \{x \mid x \in \textbf{x}_i \cap \textbf{x}_j, i \in E, j \notin E\}$. Correspondingly, its maximum boundary query is defined as\footnote{The semi-join result of $R_i \ltimes t$ is defined as function $R'_i: \D_i \to \mathbb{Z}^+$ such that $R_t(t') = R(t')$ if $\pi_{\x_i} t' = \pi_{\x_i} t$ and $0$ otherwise.}:
\begin{equation}
\label{eq:TE}
    T_E(\I) = \max_{t \in \dom(\partial E)} \sum_{t' \in \dom(\cup_{i\in E}\x_i): \pi_{\partial E} t' = t} \prod_{i \in E} R^\I_i(\pi_{\x_i} t').
\end{equation}
 
\begin{definition}[Residual Sensitivity~\cite{dong2021residual, dong2021nearly}] 
 \label{def:residual}
 Given $\beta > 0$, the \emph{$\beta$-residual sensitivity} of $\countsize(\cdot)$ over $\I$ is defined as 
 \[\RS^\beta_\countsize(\I) = \max_{k\ge 0} e^{-\beta k} \cdot \hat{\LS}^{k}_\countsize(\I),\] 
 where
 $\displaystyle{\hat{\LS}^{k}_\countsize(\I) =  \max_{\s \in \mathcal{S}^k} \max_{i \in [m]} \sum_{E \subseteq [m] \smallsetminus\{i\} }T_{[m] \smallsetminus \{i\} \smallsetminus E}(\I) \cdot \prod_{j \in E} \s_j}$,
    for $\mathcal{S}^k = \left\{\s = (\s_1, \dots, \s_m): \sum_{i=1}^m \s_i = k, \s_i \in \mathbb{Z}^{\ge0}, \forall i \in [m]\right\}$. 
 \end{definition}
 
It is well-known that $\RS^\beta_\countsize(\cdot)$ is a smooth upper bound on $\LS_\countsize(\cdot)$~\cite{dong2021residual}. However, we do not use it (as in~\cite{dong2021residual}) to 
calibrate the noise drawn from  a Laplace or Cauchy distribution. In our data release problem, we always find an upper bound $\tDelta$ for $\RS^\beta_\countsize(\I)$, which is then passed to the single-table PMW algorithm. To compute $\tDelta$, we observe that $\ln\left(\RS^\beta_\countsize(\cdot)\right)$ has global sensitivity at most $\beta$. Therefore, adding to it an appropriately calibrated (truncated and shifted) Laplace noise provides an upper bound that is private. The idea is formalized in \Cref{alg:multi-table}. Its privacy guarantee is immediate:

\begin{lemma}
\label{lem:multi-table-DP}
    Algorithm~\ref{alg:multi-table} is $(\epsilon,  \delta)$-DP.
\end{lemma}


\paragraph{Error analysis.} 
By definition of $\TLap$, we have $\left|\TLap_{2\beta/\eps}^{\tau(\eps/2,\delta/2,\beta)}\right| \leq 2\tau(\eps/2,\delta/2,\beta) \leq O\left(\frac{\beta}{\eps} \cdot \log(1/\delta)\right) = O(1)$ and therefore $\tDelta$ is a constant-approximation of $\RS^\beta_\countsize(\I)$, i.e., $\tDelta =\Theta(\RS^\beta_\countsize(\I))$. 
Hence $\widehat{n} = O\left(\countsize(\I) + \RS^\beta_\countsize(\I)\cdot \lambda \right)$. Putting everything together, the total error (omitting $\fup$) is: 
\[O\left(\sqrt{\widehat{n}} \cdot \sqrt{\tDelta} \right) = O\left(\sqrt{\countsize(\I) \cdot \RS^\beta_\countsize(\I)} + \RS^\beta_\countsize(\I) \cdot \sqrt{\lambda} \right).\]

Extending the previous lower bound argument on the two-table query, we can obtain the lower bound on the multi-table query in Theorem~\ref{the:lb-multi-table}. Moreover, we give the worst-case analysis on the error achieved above in Appendix~\ref{appendix:join-as-one}.

%% file: uniformize.tex
\section{Uniformized Sensitivity} 
\label{sec:uniformization}

So far, we have shown a parameterized algorithm for answering linear queries whose utility is in terms of the join size and the residual sensitivity. A natural question arises: can we achieve a more fine-grained parameterized algorithm with better utility? 

\begin{figure}
    \centering
     \includegraphics[scale=0.7]{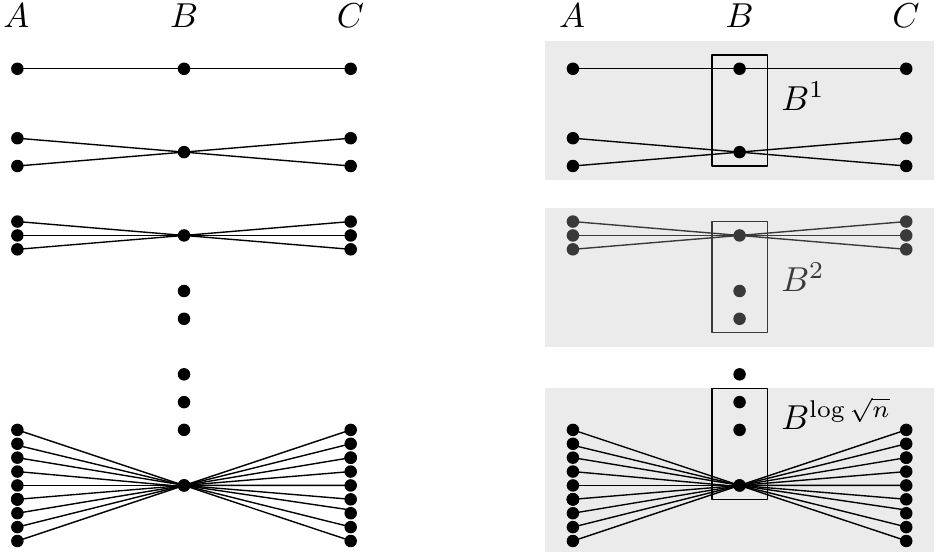}
     \caption{An instance beyond the scope of Theorem~\ref{the:lb-two-table-2}. There are $\sqrt{n}$ join values in attribute $B$, where there is exactly one join value with degree $i$ in both $R_1, R_2$ for every $i\in[\sqrt{n}]$.}
    \label{fig:hard-two-table}
\end{figure}

Let us start with an instance of two-table join (see Figure~\ref{fig:hard-two-table}), with input size $\Theta(n)$, join size $\Theta(n\sqrt{n})$, and local sensitivity $\sqrt{n}$. Algorithm~\ref{alg:two-table} achieves an error of $O(n)$. However, this instance is beyond the scope of Theorem~\ref{the:lb-two-table-2}, as the degree distribution over join values is extremely non-uniform. 
Revisiting the error bound in Theorem~\ref{the:up-2table}, we can gain an intuition regarding why Algorithm~\ref{alg:two-table} does not perform well on this instance. The costly term is $O(\sqrt{\countsize(\I)} \cdot \sqrt{\Delta})$, where $\Delta$ is the largest degree of join values in the input instance $\I$. However, there are many join values whose degree is much smaller than $\Delta$.  Therefore, a natural idea is to {\em uniformize sensitivities}, i.e., partition the input instance into a set of sub-instances by join values, where join values with similar sensitivities are in the same sub-instance. We then invoke our previous join-as-one algorithm as a primitive on each sub-instance independently, and return the union of the synthetic datasets generated for all sub-instances.  Our uniformization framework is illustrated in Algorithm~\ref{alg:uniformization-framework}. (All missing proofs are given in Appendix~\ref{appendix:uniformization}).

\subsection{Uniformized Two-Table Join}
\label{sec:two-table-uniformize}
As mentioned, there is quite an intuitive way to uniformize a two-table join.
As described in Algorithm~\ref{alg:partition-2table}, the high-level idea is to bucket join values in $\dom(B)$ by their maximum degree in $R_1$ and $R_2$. To protect the privacy of individual tuples, we draw a random sample from  $\TLap_{2/\eps}^{\tau(\eps/2,\delta/2,1)}$ and add it to a join value's degree, before determining to 
which bucket it should go. Recall that $\lambda = \frac{1}{\epsilon} \log \frac{1}{\delta}$. Let $\gamma_0 = 0$ and $\gamma_i = 2^i \lambda$ for all $i \in \N$. Conceptually, we divide $[0, n+ \lambda]$ into $\ell = \lceil \log (\frac{n}{\lambda}+1) \rceil$ buckets, where the $i$th bucket is associated with $(\gamma_{i - 1}, \gamma_i]$ for $i \in [\ell]$. The set of values from $\dom(B)$ whose maximum noisy degree in $R_1$ and $R_2$ falls into bucket $i$ is denoted as $B^{i}$. For each $i$, we identify tuples in $R_1, R_2$ whose join value falls into $B^{i}$ as $R^{i}_1, R^{i}_2$, which forms the sub-instance $(R^i_1, R^i_2)$. More specifically, $R^i_1: B^i \to \mathbb{Z}^{\ge0}$, such that $R^i_1(t) = R_1(t)$ if $\pi_{B} t \in B^i$ and $R^i_1(t) = R_1(t)$ otherwise. $R^i_2$ is defined similarly.  Finally, we 
return all the sub-instances as the partition.
\begin{algorithm}[t]
\caption{{\sc Uniformize}$_{\eps.\delta}(\I)$}
\label{alg:uniformization-framework}

$\mathbb{I} \gets \textsc{Partition}_{\eps/2, \delta/2}(\I)$\;
\ForEach{$\I' \in \mathbb{I}$}{
    $\F(\I') \gets \textsc{MultiTable}_{\eps/2,\delta/2}(\I')$; \hfill 
 $\blacktriangleright$\Cref{alg:multi-table}\;}
\Return $\bigcup_{\I' \in \mathbb{I}}\F(\I')$\;
\end{algorithm}
\begin{algorithm}[t]
\caption{{\sc Partition-TwoTable}$_{\eps, \delta}(\I=\{R_1, R_2\})$}
\label{alg:partition-2table}

\lForEach{$i \in \N$}{$B^i \gets \emptyset$}
\ForEach{$b \in \dom(B)$}{
   $\tdeg_B(b) = \max\{\deg_{1,B}(b), \deg_{2,B}(b)\} + \TLap_{1/\eps}^{\tau(\eps,\delta,1)}$\;
    $i \gets \max\left\{1, \left\lceil \log \frac{1}{\lambda} \cdot \tdeg_B(b) \right\rceil\right\}$\;
    $B^{i} \gets B^{i} \cup \{b\}$\;
}
\ForEach{$i$ with $B^i \neq \emptyset$}{
    \ForEach{$j\in \{1,2\}$}{
    $R^{i}_j: \dom(\mathbb{D}_1) \to \mathbb{Z}^{\ge0}$ such that for any $t \in \mathbb{D}_1$, $R^{i}_j(t)= R_j(t)$ if $\pi_B t \in B^i$ and $R^{i}_j(t)= 0$ otherwise\;}
}
\Return $\bigcup_{i: B^i \neq \emptyset} \{(R^i_1, R^i_2)\}$\;
\end{algorithm}

%
%

\begin{lemma}
\label{lem:partition-2table-DP}
    Algorithm~\ref{alg:uniformization-framework} on two-table join is $(\epsilon, \delta)$-DP.
\end{lemma}

The key insight in Lemma~\ref{lem:partition-2table-DP} is that adding or removing one input tuple can increase or decrease the degree of one join value $b \in \dom(B)$ by at most one. Hence, Algorithm~\ref{alg:partition-2table} satisfies $(\epsilon,\delta)$-DP by  parallel composition~\cite{mcsherry2009privacy}. Moreover, since each input tuple participates in exactly one sub-instance, and Algorithm~\ref{alg:pmw} preserves $(\epsilon, \delta)$-DP for each sub-instance by Lemma~\ref{lem:two-table-DP}, Algorithm~\ref{alg:uniformization-framework} preserves $(2\epsilon, 2\delta)$-DP by basic composition~\cite{dwork2006calibrating}.

\paragraph{Error Analysis.} 
Note that $n$ is not explicitly used in Algorithm~\ref{alg:partition-2table} as this is not public, but it is easy to check that there exists no join value in $\dom(B)$ with degree larger than $n$ under the input size constraint. Hence, it is safe to consider $i \in [\ell]$ in our analysis.

Given an instance $\I$ over the two-table join, let $\pi = \{B^1_\pi,\dots, B^{\ell}_\pi\}$ be the partition of $\dom(B)$ generated by Algorithm~\ref{alg:partition-2table}. 
Let $\I^i_\pi = (R^i_1, R^i_2)$ be the sub-instance induced by $B^i_\pi$.  
Let $\F^i$ be the synthetic dataset generated for $\I^i_\pi$. From Theorem~\ref{the:up-2table}, with probability $1-1/\textsf{poly}(|\Q|)$, the error for answering any linear query defined on $\left(\dom(A) \times \dom(B^i_\pi)\right) \times \left(\dom(B^i_\pi) \times \dom(C)\right)$ with $\F^i$ is (omitting $\fup$) $\alpha_i = O\left(\sqrt{\countsize(\I^i_\pi) \cdot 2^i \cdot \lambda} + 2^i\cdot \lambda^{3/2}\right)$. 
By a union bound, with probability $1- 1/\poly(|\Q|)$, the error for answering any linear query in $\Q$ with $\cup_i \F^i$ is (omitting $\fup$):
\begin{equation}
\label{eq:error-2table-uniform}
    \alpha \le \sum_{i\in [\ell]} \alpha_i \le  \lambda^{3/2} \cdot (\Delta+\lambda) + \sqrt{\lambda} \cdot \sum_{i\in [\ell]} \sqrt{\countsize(\I^i_\pi)} \cdot \sqrt{2^i},
\end{equation}
since $\sum_{i\in [\ell]} 2^i \le  \sum_{i\in [\lceil\log (\Delta+\lambda)\rceil]} 2^i=O \left(\Delta + \lambda\right)$. 
Moreover, 
we observe that Algorithm~\ref{alg:uniformization-framework} achieves better (or at least not worse than) error than Algorithm~\ref{alg:two-table} (if ignoring $\lambda$), since 
\begin{align*}
    (\ref{eq:error-2table-uniform}) &\le \lambda^{3/2} \cdot (\Delta + \lambda) + \sqrt{\lambda} \cdot \sqrt{\sum_{i\in [\ell]} \countsize(\I^i_\pi)} \cdot \sqrt{\sum_{i\in [\ell]} 2^i} \\
    & = \lambda^{3/2} \cdot (\Delta + \lambda)  + \sqrt{\lambda} \cdot \sqrt{\countsize(\I)} \cdot \sqrt{\Delta + \lambda},
\end{align*}
where the first inequality is implied by the Cauchy--Schwarz inequality. Furthermore, we observe that the gap between the error achieved by Algorithm~\ref{alg:uniformization-framework} and Algorithm~\ref{alg:two-table} can be 
polynomially large in terms of the data size; see Example~\ref{exp:gap}.

\begin{example}
\label{exp:gap}
    Consider an instance $\I$ of two-table join, which further amplifies the non-uniformity of the instance in Figure~\ref{fig:hard-two-table}. For $i \in \{0,1,\cdots, \frac{2}{3}\log_2 k\}$, there are $k^2/8^i$ distinct join values in $\dom(B)$ with $\deg_{1,B}(b) = \deg_{2,B}(b) = 2^i$. Obviously, $\Delta = k^{2/3}$. It can be easily checked that the input size is $n \le 2k^2$ and join size is $\countsize(\I) = \frac{2}{3} k^2\log_2 k$. For simplicity, we assume $\epsilon = \Theta(1)$ and $\delta = 1/n^c$ for some constant $c$, therefore $\lambda = \Theta(1)$. 
    Algorithm~\ref{alg:two-table} achieves an error of (omitting $\fup$)
    $\alpha = O(\sqrt{\countsize(\I) \cdot \Delta}) = O(k^{4/3})$. By uniformization, join values with the same degree will be put into the same bucket (even after adding a small noise). In this way, \Cref{alg:partition-2table} achieves an error of (omitting $\fup$):
    $\displaystyle{\alpha = O(k^{2/3} + \sum_{i\in [\frac{2}{3}\log_2 k]} \sqrt{k^2/8^i \cdot 2^i \cdot 2^i} \cdot \sqrt{2^i})= O(k \log_2 k)}$,
    improving Algorithm~\ref{alg:two-table} by a factor of $k^{1/3} = O(n^{1/6})$.
\end{example}

Although the partition generated by Algorithm~\ref{alg:partition-2table} is randomized due to noisy degrees (line 3), it is not far away from a fixed partition based on true degrees.  As shown in Appendix~\ref{appendix:uniformization}, Algorithm~\ref{alg:partition-2table} can have its error bounded by what can be achieved through the {\em uniform partition} below (recall that $\ell = \lceil \log \frac{n}{\lambda} \rceil$ and $\gamma_i = \lambda \cdot 2^i$ for $i \in [\ell]$):

\begin{definition}[Uniform Partition]
    For an instance $\I = (R_1, R_2)$, a uniform partition of $\dom(B)$ is $\pi^* = \left\{B^1_{\pi^*}, \dots, B^{\ell}_{\pi^*}\right\}$ such that for any $i \in [\ell]$, $b \in B^i_{\pi^*}$ if
    $\displaystyle{\max\left\{\deg_{1,B}(b), \deg_{2,B}(b)\right\} \in (\gamma_{i-1}, \gamma_i]}$.
\end{definition}

\begin{theorem}
\label{the:up-2table-uniformize}
    For any two-table instance $\I$, a family $\Q$ of linear queries, and $\epsilon > 0$, $\delta > 0$, there exists an algorithm that is $(\epsilon, \delta)$-DP, and with probability at least $1- 1/\textsf{poly}(|\Q|)$ produces $\F$ such that all linear queries in $\Q$ can be answered to within error: 
    \[\alpha = O\left((\lambda^{\frac{3}{2}} (\Delta + \lambda) + \sum_{i\in [\ell]} \sqrt{\countsize\left(\I^i_{\pi^*}\right)} \cdot \sqrt{2^i \cdot \lambda}) \cdot \fup \right), \]
    where $\ell = \lceil \log \frac{n}{\lambda} \rceil$ and $\I^i_{\pi^*}$ is the sub-instance of $\I$ induced by $B^i_{\pi^*}$.
\end{theorem}

Let $\overrightarrow{\OUT} = \left \langle \OUT^i \in \mathbb{Z}^{\ge0}:i \in [\ell], \sum_i \OUT^i \in [0,n^2]\right\rangle$ be a join size vector. An instance $\I=(R_1, R_2)$ {\em conforms} to $\overrightarrow{\OUT}$ if
\begin{itemize}[leftmargin=*]
    \item for every $i \in [\ell]$, $\countsize\left(\I^i_{\pi^*}\right) = \Theta\left(\OUT^i\right)$;
    \item $\countsize(\I) = \Theta\left(\sum_{i \in [\ell]} \OUT^i\right)$.
\end{itemize}
Then, we introduce a more fine-grained lower bound parameterized by the join size distribution under the uniform partition, and show that Algorithm~\ref{alg:uniformization-framework} is optimal  up to poly-logarithmic factors. The proof is given in Appendix~\ref{appendix:uniformization}.

\ifbeforepasinchange
\else


\begin{theorem}
\label{the:lb-instance-optimality-2table}
    Given a join size vector $\overrightarrow{\OUT}$,
    for every sufficiently small $\eps >0$, $n_D \geq (\log \OUT )^{O(1)}$ and $n_Q \geq (\OUT \cdot \log n_D)^{O(1)}$, there exists a family $\Q$ of queries of size $n_Q$ on a domain $\D$ of size $n_D$ such that any $(\epsilon, o(1/n))$-DP algorithm that takes as input a two-table instance of input size at most $n$ while conforming to $\overrightarrow{\OUT}$, and outputs an approximate answer to each query in $\Q$ to within error $\alpha$ must satisfy  
    $\displaystyle{\alpha \ge \tOmega\left(\max_{i\in [\ell]} \min\left\{\OUT^i, \sqrt{\OUT^i} \cdot \sqrt{2^i \cdot \lambda} \cdot \flo\right\}\right)},$ for $\ell = \lceil \log \frac{n}{\lambda} \rceil$.
\end{theorem}

\fi


\input{hierarchical}

%% file: hierarchical.tex
\subsection{Uniformized Hierarchical Join} 
\label{sec:hierarchical}

\begin{figure}[t]
    \centering
    \includegraphics[scale=0.58]{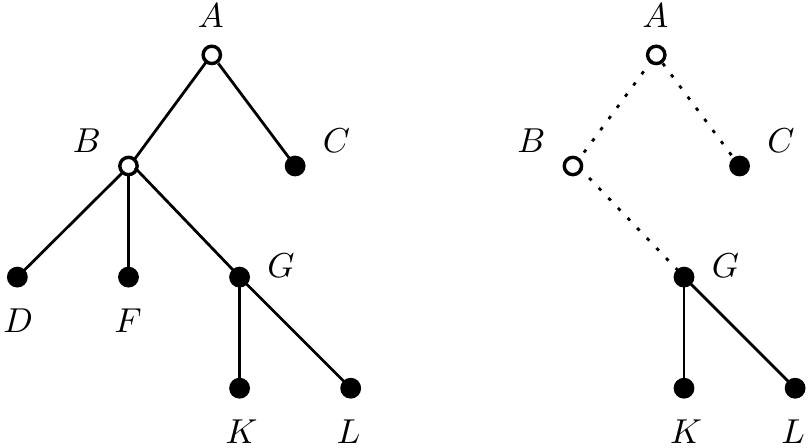}
    \caption{Left is the attribute tree for hierarchical join $\mathcal{H}$ with $\x = \{A,B,C,D,F,G,K,L\}$, $\x_1 =\{A,B,D\}$, $\x_2 = \{A,B,F\}$, $\x_3 =\{A,B,G,K\}$, $\x_4 = \{A,B,G,L\}$, and $\x_5 = \{A,C\}$. Right is the residual query defined on $E = \{3,4,5\}$ with $\partial E = \{A,B\}$. Moreover, $\bigwedge_{345} = \{A\}$ and $\bigvee_{345} = \{A,B,C,G,K,L\}$. $\mathcal{H}_{E, \partial E}$ is disconnected with $\mathcal{C}_E = \{\{3,4\}, \{5\}\}$. $T_{345}(\I)$ can be upper bounded by $\degg_{5}(A) \cdot \degg_{34}(AB) \cdot \degg_{3}(ABG) \cdot \degg_{4}(ABG)$.}
    \label{fig:hierarchical}
    \end{figure}

This uniformization technique, surprisingly, can be further extended beyond two-table queries to the class of {\em hierarchical queries}. A join query $\H = \{\x, \{\x_1, \dots, \x_m\}\}$ is {\em hierarchical}, if for any pair $x,y$ of attributes, either $\atom(x) \subseteq \atom(y)$, or $\atom(y) \subseteq \atom(x)$, or $\atom(x) \cap \atom(y) = \emptyset$, where $\atom(x) = \{i \in [m]: x \in \x_i\}$ is the set of relations containing attribute $x$. One can always organize the attributes of a hierarchical join into a tree such that every relation corresponds to a root-to-node path (see Figure~\ref{fig:hierarchical}). 

We show how to exploit this nice property to improve Algorithm~\ref{alg:multi-table} on hierarchical joins with uniformization. Recall that the essence of uniformization is to decompose the instance into a set of sub-instances so that we can upper bound the residual sensitivity as tight as possible, as implied by the error expression in Theorem~\ref{the:ub-multi-table}. In Definition~\ref{def:residual}, the residual sensitivity is built on the join sizes of a set of maximum boundary queries $T_E(\I)$'s, but these statistics are too far away from being the partition criteria. Instead, we find an upper bound of residual sensitivity in terms of degrees. 

\subsubsection{{\bf An Upper Bound on $T_E$}} 
In a hierarchical join $\mathcal{H}$, we denote $\mathcal{T}$ as the attribute tree for $\mathcal{H}$. For any $E \subseteq [m]$, the partial attributes $\partial E$ forms a connected subtree of $\mathcal{T}$ including the root (see Figure~\ref{fig:hierarchical}). 
To derive an upper bound on $T_E$, we identify a broader class of {\em $q$-aggregate queries} by generalizing aggregate attributes from $\partial_E$ to any subset of attributes that form a connected subtree of $\mathcal{T}$ including the root (i.e., sitting on top of $\mathcal{T}$). For simplicity, we define $\bigwedge_E = \cap_{i \in E}\x_i$ and $\bigvee_E = \cup_{i\in E} \x_i$.

\begin{definition}[q-aggregate Query]
\label{def:bar-q}
    For a hierarchical join $\H = (\x, \{\x_1, \dots, \x_m\})$, an instance $\I$, and a subset $E \subseteq [m]$ of relations, assume that a subset $\y \subseteq \bigvee_E$ of attributes satisfies the following property\footnote{The same property on $\y$ has been characterized by {\em q-hierarchical queries} \cite{berkholz2017answering}, where $\y$ serves as the set of output attributes there.}: for any $x_1, x_2 \in \x$, if $\atom(x_1) \subseteq \atom(x_2)$ and $x_1 \in \y$, then $x_2 \in \y$.
    A \emph{$q$-aggregate query} defined over $E$ on $\y$ is defined as
    \[T_{E,\y}(\I) = \max_{t \in \dom(\y)} \sum_{t' \in \dom(\bigvee_E): \pi_{\y} t'= t} \prod_{i \in E} R_i(\pi_{\x_i} t').\] 
\end{definition}

It is not hard to see $T_E(\I) = T_{E, \partial E}(\I)$ from (\ref{eq:TE}). Below, we focus on an upper bound for $T_{E,\y}(\I)$ with general $\y$ characterized by Definition~\ref{def:bar-q}, which boils down to the product of {\em maximum degrees}:
\begin{definition}[Maximum Degree]
\label{def:degree-function}
    For a hierarchical join $\H = (\x, \{\x_1, \dots, \x_m\})$, an instance $\I$, $E \subseteq [m]$ and $\y \subseteq \bigwedge_E$, the degree of tuple $t \in \dom(\y)$ is:
    \begin{displaymath}
    \deg^\I_{E,\y}(t) = \left\{ \begin{array}{ll}
    \displaystyle{\sum_{t' \in \dom(\x_i): \pi_{\y} t' = t} R_i(t')} &\textrm{if $|E| =1$, say $E = \{i\}$}\\
    \displaystyle{\left|\{t' \in \Psi_E(\I): \pi_{\y} t' =t\}\right|} &\textrm{otherwise,}
    \end{array} \right.
    \end{displaymath}
    where 
    $\Psi_E(\I) = \left\{\pi_{\bigwedge_E} t':t' \in \dom(\bigvee_E), \prod_{i \in E} R_i(\pi_{\x_i} t') > 0\right\}$. The \emph{maximum degree}
    is defined as
    $\displaystyle{\degg^\I_{E}(\y) = \max_{t \in \dom(\y)} \deg^\I_{E,\y}(t)}$.
\end{definition}
We next show an upper bound on $T_{E,\y}(\I)$ using $\degg^\I_E(\y)$'s. When the context is clear, we drop the superscript $\I$ from $\degg^\I_E(\y)$.
{\bf Case (1).} When $|E|=1$, say $E = \{i\}$, we note that $\y \subseteq \x_i$ and $T_{E,\y}(\I)$ is essentially equivalent to $\degg_{i}(\y)$, since 
\begin{align*}
T_{E,\y}(\I) &= \max_{t \in \dom(\y)}\sum_{t' \in \dom(\x_i): \pi_{\y} t' = t} R_i(t') = \degg_E(\y).
\end{align*}

{\bf Case (2).} In general, let $\mathcal{H}_{E,\y} = (\bigvee_E - \y, \{\x_i - \y: i \in E\})$ be the residual join defined on relations in $E$ after removing attributes $\y$. We next distinguish two cases based on the connectivity\footnote{A multi-way join $\mathcal{H} = (\x, \{\x_i:i\in [m]\})$ can be modeled it as a graph $G_\mathcal{H}$, where each $\x_i$ is a vertex and an edge exists between $\x_i, \x_j$ if $\x_i \cap \x_j \neq \emptyset$. $\mathcal{H}$ is {\em connected} if $G_\mathcal{H}$ is connected, and {\em disconnected} otherwise. For disconnected $\mathcal{H}$, we can decompose it into multiple connected subqueries by finding all connected components for $G_\mathcal{H}$ via graph search algorithm, where each component indicates a connected subquery of $\mathcal{H}$.} of $\mathcal{H}_{E,\y}$:

{\bf Case (2.1): $\mathcal{H}_{E,\y}$ is disconnected.} Let $\mathcal{C}_E$ be the set of connected subqueries of $\mathcal{H}_{E,\y}$. We can further decompose $ T_{E,\y}(\I)$ as:
    \begin{align*}
        T_{E,\y}(\I) = & \max_{t \in \dom(\y)} \prod_{E' \in \mathcal{C}_E} \sum_{t' \in \dom(\bigvee_{E'}): \pi_{\y \cap (\bigvee_{E'})} t' = t} \prod_{i \in E'} R_i(\pi_{\x_i} t') \ \ \ \ \ \ \ \\
        \le & \prod_{E' \in \mathcal{C}_E} \max_{t \in \dom(\y)} \sum_{t' \in \dom(\bigvee_{E'}): \pi_{\y \cap (\bigvee_{E'})} t' = t} \prod_{i \in E'} R_i(\pi_{\x_i} t') \\
        =& \prod_{E' \in \mathcal{C}_E} T_{E', \y \cap (\bigvee_{E'})}(\I),
    \end{align*}
    since $\y \cap (\bigvee_{E'}) \subseteq \bigvee_{E'}$ and $\y \cap (\bigvee_{E'})$ satisfies the property in Definition~\ref{def:degree-function}, if $\y$ satisfies the same property.  

    {\bf Case (2.2): $\mathcal{H}_{E,\y}$ is connected.} In this case, we have $\y \subsetneq \bigwedge_E$.
    \begin{align*}
        T_{E,\y}(\I) 
        \le \ & \degg_{E}(\y) \cdot \max_{t \in \dom(\bigwedge_E)} \sum_{t' \in \dom(\bigvee_E): \pi_{\bigwedge_E} t'= t} \prod_{i \in E} R_i(\pi_{\x_i} t') \\
        = \ & 
        \degg_{E}(\y) \cdot T_{E, \bigwedge_E}(\I),
    \end{align*}
    since $\bigwedge_E \subseteq \bigvee_E$ and $\bigwedge_E$ satisfies the property in Definition~\ref{def:degree-function}.    

   Hence, $T_{E,\y}(\I)$ is eventually upper bounded by a product chain of maximum degrees (see Figure~\ref{fig:hierarchical}). A careful inspection reveals that the maximum degrees participating in $T_{E}(\I)$ for $\y = \partial E$ are not arbitrary; instead, they display rather special structures captured by Lemma~\ref{lem:hierarchical-degree-function}, which is critical to our partition procedure.


\begin{lemma}
\label{lem:hierarchical-degree-function}
    Every maximum degree $\degg_{E'}(\y)$ participating in the upper bound of $T_{E}(\I)$ corresponds to a distinct attribute $x \in \x$ such that $E' = \textrm{atom}(x)$ and $\y$ corresponds to the ancestors of $x$ in $\mathcal{T}$. 
\end{lemma}


\subsubsection{{\bf Partition with Maximum Degrees}}

After getting an upper bound on $T_{E}(\I)$, we now define {\em degree configuration} for hierarchical joins, similarly to the two-table join. Our target is to decompose the input instance into a set of sub-instances (which may not be tuple-disjoint), such that each sub-instance is characterized by one distinct {\em degree configuration}, and join results of all sub-instances form a partition of the final join result. 

\begin{definition}[Degree Configuration]
\label{def:degree-characterization}
    For a hierarchical join $\H =(\x, \{\x_1,\dots,\x_m\})$, a \emph{degree configuration} is defined as $\sigma: 2^{[m]}\times 2^{\x} \to \mathbb{Z}^{\ge0} \cup \{\bot\}$ such that for any $E \subseteq [m]$ and $\y \subseteq \x$, $\sigma(E, \y) \neq \bot$ if and only if there exists an attribute $x \in \x$ such that $E = \atom(x)$ and $\y$ is the set of ancestors of $x$ in the attribute tree $\mathcal{T}$ of $\mathcal{H}$.
\end{definition}

Algorithm~\ref{alg:uniformize-hierarchical} recursively decomposes the input instance by attributes in a bottom-up way on $\mathcal{T}$, and invokes  Algorithm~\ref{alg:partition-hierarchical} as a primitive to further decompose every sub-instance. Algorithm~\ref{alg:partition-hierarchical} takes as input an attribute $x$ and an instance $\I$ that has been decomposed by all descendants of $x$, 
and outputs a set of sub-instances such that their join results form a partition of the join result of $\I$, and 
    for every sub-instance $\I'$, $\deg^{\I'}_{\atom(x), \y}(t)$ is roughly the same for every tuple $t \in \dom(\y)$, where $\y$ is the set of ancestors of $x$ in $\mathcal{T}$.

\begin{algorithm}[t]
\caption{{\sc Partition-Hierarchical}$_{\eps, \delta}(\mathcal{H}, \I)$}
\label{alg:uniformize-hierarchical}

$\mathbb{I} \gets \{\I\}$, $\mathcal{T} \gets$ an attribute tree of $\mathcal{H}$\;
\While{there exists a non-visited node in $\mathcal{T}$}{
    $\mathbb{I}' \gets \emptyset$\;    
    $x \gets$ a leaf or any node whose children are all visited\;
    \ForEach{$\I' \in \mathbb{I}$}{
        $\mathbb{I}' \gets \mathbb{I}' \cup \textsc{Decompose}_{\eps, \delta}(\I', x)$; \hfill $\blacktriangleright$\Cref{alg:partition-hierarchical} \;}
    $\mathbb{I}\gets \mathbb{I}'$, Mark $x$ as visited\;
}
\Return $\mathbb{I}$\;
\end{algorithm}

\begin{algorithm}[t]
\caption{{\sc Decompose}$_{\eps, \delta}(\I, x)$}
\label{alg:partition-hierarchical}

$\y \gets \{y \in \x: \atom(x) \subsetneq \atom(y)\}$, $E \gets \atom(x)$\;
\lForEach{$i \in \mathbb{N}$}{$\y^i \gets \emptyset$}
\ForEach{$t \in \dom(\y)$}{
    $\tdeg^{\I}_{E,\y}(t) = \deg^{\I}_{E,\y}(t) + \TLap_{1/\eps}^{\tau(\eps,\delta,1)}$\;
    $i \gets \max\left\{1, \left\lceil \log \frac{1}{\lambda} \cdot \tdeg^{\I}_{E,\y}(t) \right\rceil\right\}$\;
    $\y^i \gets \y^i \cup \left\{t\right\}$\;
}
\ForEach{$i$ with $\y^i \neq \emptyset$}{
    \ForEach{$j \in E$}{
    $R^{\I}_{j,i}: \dom(\mathbb{D}_j) \to \mathbb{Z}$ such that for $t \in \mathbb{D}_j$, $R^{\I}_{j,i}(t)=$ $R^{\I}_j(t)$ if $\pi_\y t \in \y^i$ and $R^{\I}_{j,i}(t)= 0$ otherwise;}
}
\Return $\bigcup_{i: \y^i \neq \emptyset} \left\{\{R^{\I}_{j,i}: j \in E\} \cup \{R^{\I}_{j}: j \notin E\} \right\}$\;
\end{algorithm}

\begin{lemma}
\label{lem:partition-hierarchical}
    For input instance $\I$, let $\mathbb{I}$ be the set of sub-instances returned by Algorithm~\ref{alg:partition-hierarchical}. $\mathbb{I}$ satisfies the following properties:
    \begin{itemize}[leftmargin=*]
        \item For any $\vec{t} \in \times_{i \in [m]}\mathbb{D}_i$, there exists some $\I' \in \mathbb{I}$ such that $\mathrm{Join}^\I\left(\vec{t}\right) =  \mathrm{Join}^{\I'}\left(\vec{t}\right)$ and  $\mathrm{Join}^{\I''}\left(\vec{t}\right) = 0$ for any $\I'' \in \mathbb{I} - \{\I'\}$.
        \item Each input tuple appears in $O\left(\log^c n\right)$ sub-instances of $\mathbb{I}$, where $c$ is a constant depending on $m$;
        \item Each sub-instance $\I \in \mathbb{I}$ corresponds to a distinct degree configuration $\sigma$ such that for any $E \in [m]$ and $\y \subseteq \x$ with $\sigma(E,\y)\neq \bot$:
        \begin{equation*}
            \label{eq:degree-sigma}
            \tdeg^{\I}_{E,\y}(t) \in \left(\lambda \cdot 2^{\sigma(E,\y)-1}, \lambda \cdot 2^{\sigma(E,\y)}\right]
        \end{equation*}
        holds for any $t \in \dom(\y)$. 
    \end{itemize}
\end{lemma}

\begin{lemma}
\label{lem:partition-hierarchical-DP}
    Algorithm~\ref{alg:uniformization-framework} is $(O(\log^c n) \cdot \epsilon, O(\log^c n) \cdot \delta)$-DP, where $c$ is a constant depending on $m$.
\end{lemma}

The logarithmic factor in Lemma~\ref{lem:partition-hierarchical-DP} arises since each input tuple participates in $O(\log^c n)$ sub-instances, implied by Lemma~\ref{lem:partition-hierarchical}. We present the error analysis of Algorithm~\ref{alg:uniformization-framework} for hierarchical joins and a parameterized lower bound in Appendix~\ref{appendix:uniformization}.

%% file: conclusion.tex
\section{Conclusions and Discussion}

In this paper, we proposed algorithms for releasing synthetic data for answering linear queries over multi-table joins.  Our work opens up several interesting directions listed below.

\paragraph{Non-Hierarchical Queries.} 
Perhaps the most immediate question is if uniformization can benefit the non-hierarchical case, even for the simplest join $\mathcal{H}$ defined on $\x = \{A,B,C\}$ with $\x_1 = \{A,B\}$, $\x_2 = \{B,C\}$, and $\x_3 =\{C,D\}$. In determining the residual sensitivity in Definition~\ref{def:residual}, we observe that $T_{23}(\I) \le \degg_{2}(B) \cdot \degg_{3}(C)$, $T_{12}(\I) \le \degg_{1}(B) \cdot \degg_{2}(C)$, and $T_{13} \le \degg_{1}(B) \cdot \degg_3(C)$. It is easy to uniformize $\degg_1(B), \degg_3(C)$ by partitioning $R_1, R_3$ by attributes $B,C$ respectively. However, it is challenging to uniformize $\degg_{2}(B), \degg_{2}(C)$ by partitioning $R_2$, while keeping the number of sub-instances small. For example, a trivial strategy simply puts every individual tuple $t \in \mathbb{D}_2$ with $R_2(t) >0$, together with  tuples in $R_1, R_3$ that can be joined with $t$, as one sub-instance. In this case, it is great to have small $\degg_{2}(B)$ and $\degg_{2}(C)$, but each tuple from $R_1$ or $R_3$ may participate in $\degg_1(B)$ or $\degg_3(C)$ sub-instances, hence the privacy consumption increases linearly when applying the parallel decomposition! Alternatively, one may uniformize $\degg_{2}(B)$ and $\degg_{2}(C)$ independently, say with partitions $\pi_1$ of $\dom(B)$ and $\pi_2$ of $\dom(C)$. However, the degrees $\deg_{2,B}$ as well as $\deg_{2, C}$ are defined on the whole relation $R_2$, hence we may still end up with very non-uniform distribution of $\deg_{2,B}, \deg_{2,C}$ when restricting to a sub-instance induced by $B^i_{\pi_1}, C^j_{\pi_2}$ together. 
We leave this interesting question for  future work.

\paragraph{Query-Specific Optimality.} 
Throughout this work, we consider \emph{worst-case} set $\Q$ of queries parameterized by its size. Although this is a reasonable starting point, it is also plausible to hope for an algorithm that is nearly optimal for \emph{all} query sets $\Q$. In the single-table case, this has been achieved in~\cite{HardtT10,BhaskaraDKT12,NikolovT016}. However, the situation is more complicated in the multi-table setting since we have to take the local sensitivity into account, whereas, in the single-table case, the query family already dictates the possible change resulting from moving to a neighboring dataset. This is also an interesting open question for future research.

\paragraph{Instance-Specific Optimality.}
We have considered worst-case instances in this work. One might instead prefer to achieve finer-grained instance-optimal errors. For the single-table case, \cite{asi2020instance} observed that an instance-optimal DP algorithm is not achievable, since a trivial algorithm can return the same answer(s) for all input instances, which could work perfectly on one specific instance but poorly on all remaining instances. To overcome this, a notion of ``neighborhood optimality'' has been proposed~\cite{dong2021nearly, asi2020instance}, where we consider not only a single instance but also its neighbors at some constant distance. We note, however, that this would not work in our setting when there are a large number of queries. Specifically, if we again consider an algorithm that always returns the true answer for this instance, then its error with respect to the entire neighborhood set is still quite small---at most the distance times the maximum local sensitivity in the set. This is independent of the table size $n$, whereas our lower bounds show that the dependence on $n$ is inevitable. As such, the question of how to define and prove instance-optimal errors remains open for the multi-query case.

 

%% file: appendix.tex
\appendix
\newcommand{\B}{\mathbb{J}}
\newcommand{\A}{\mathbb{F}}

\section{Guarantees of PMW}
\label{appendix:single-table}

In this section, we prove the guarantees of the single-table PMW algorithm (\Cref{alg:pmw}). We stress that this is essentially the same as the proof in~\cite{hardt2012simple}, but we reprove it for completeness.

\begin{theorem} \label{thm:pmw}
Let $\I, \I'$ be neighboring instances such that $|\countsize(\I) - \countsize(\I')| \leq \Delta$. Then, \Cref{alg:pmw} satisfies the following:
\begin{itemize}[leftmargin=*]
\item (Privacy) $\PMW(\I) \approx_{(\eps, \delta)} \PMW(\I')$.
\item (Utility) With probability at least $1 - 1/\poly(|\Q|)$, $\PMW(\I)$ produces a dataset such that all queries in $\Q$ can be answered to within error $\displaystyle{\alpha =O\left((\sqrt{\countsize(\I) \cdot \tDelta} + \tDelta \cdot \sqrt{\lambda}) \cdot \fup\right)}$.
\end{itemize}
\end{theorem}

\paragraph{Privacy Guarantee.} Let $\I = (R_1, \dots, R_m)$ and $\I'=(R'_1, \dots, R'_m)$, with $|\countsize(\I) - \countsize(\I')| \leq \Delta$. By the privacy guarantee of the truncated Laplace mechanism, we have $\hn + \TLap_{2\tDelta/\eps}^{\tau(\eps/2,\delta/2,\tDelta)} \approx_{(\eps/2, \delta/2)} \hn' + \TLap_{2\tDelta/\eps}^{\tau(\eps/2,\delta/2,\tDelta)}$. Let us now condition on $\hn = \hn'$. Let $\F, \F'$ be the synthetic data generated for $\I, \I'$ correspondingly. The algorithm starts with the same uniform distribution. Furthermore, in each update, the guarantees of the exponential mechanism (EM) and the Laplace mechanism ensure that (conditioned on results from iteration $1, \dots, i - 1$ being the same), we have $(i, m_i) \approx_{2\eps'} (i', m'_{i'})$. By applying advanced composition, we can conclude that, conditioned on $\hn = \hn'$, we have that $\F \approx_{(\eps/2,\delta/2)} \F'$. Finally, by basic composition (over the $\hn'$ part and the computation of $\F$ part), we can conclude that $\F \approx_{(\eps,\delta)} \F'$.

\paragraph{Utility Guarantee.}
For convenience, let $\B$ denote the join result of $\I$, and let $n = |\countsize(\I)|$. Define $\ME_i = \max_q |q(\A_{i-1}) - q(\B)|$ as the maximum error over all queries in terms of $\A_{i-1}$ and $\B$. Now we are going to bound the maximum error of the PMW algorithm:
\begin{align*}
   \max_q |q(\textrm{avg}_{i \le k} \A_i) - q(\B)|
    & = \max_q |\textrm{avg}_{i \le k} q(\A_i) - q(\B)|,
\end{align*}
which can be further upper bounded by 
\[ \le \max_q \textrm{avg}_{i \le k} |q(\A_i) - q(\B)| \le \textrm{avg}_{i \le k} \max_q |q(\A_i) - q(\B)| \le \textrm{avg}_{i \le k} \ME_i.\]
We next state the guarantees from the exponential mechanism and the Laplace mechanism. Let $\AE = \tDelta \cdot \log |\Q| / \eps'$.
\begin{lemma}
\label{lem:error}
With probability at least $1-2k/|\Q|^c$ for any $c \ge 0$, for all $1 \le i \le k$, we have:
\begin{align*}
    & |q_i(\A_{i-1}) - q_i(\B)| \ge \ME_i - (2c + 2) \times \AE, \textit{and } |m_i - q_i(\B)| \le c \times \AE.
\end{align*}
\end{lemma}

\begin{proof}
For the first inequality, we note that the probability EM with parameter $\eps'$ and sensitivity $\tDelta$ selects a query with quality score at least $r$ less than the optimal is bounded by 
\[\Pr[|q_i(\A_{i-1}) - q_i(\B)| < \ME_i - (2c+2) / \eps'] \le |\Q| \times \exp\left(-\frac{\gamma \cdot \eps'}{\tDelta}\right) \le \frac{1}{|\Q|^c}. \]
For the second inequality, we note that $|m_i - q_i(\B)|  \le c \cdot \gamma$ if and only if $\left|\textsf{Lap}_{\tDelta/\eps'}\right| \le c \cdot \gamma$. From Laplace distribution, we have:
\begin{align*}
    & \Pr\left[\left|\textsf{Lap}_{\tDelta/\eps'}\right| > c \cdot (\tDelta/\eps') \cdot  \log |\Q| \right] \le \exp(-c \cdot \log |\Q|) = 1/|\Q|^c.
\end{align*}
A union bound over $2k$ events completes the proof.
\end{proof}
We consider how the PMW mechanism can improve the approximation in each round where $q_i(\A_{i}) - q_i(\B)$ has large magnitude. To capture the improvement, we use the relative entropy function:
\begin{align*}
\Psi'_i = \frac{1}{n} \sum_{x \in \D} \B(x) \log \left(\frac{\B(x)}{\A_i(x)}\right), \ \Psi_i = \frac{n}{\hn} \cdot \Psi_i.
\end{align*}
Here, we can only show that $\Psi_0 \le \log |\D|$ and $\Psi_i \ge -1$.

We next consider how $\Psi$ changes in each iteration:
\begin{align*}
    \Psi_i - \Psi_{i-1} = & \frac{1}{\hn} \cdot \sum_{x \in \D} \B(x) \cdot\log \left(\frac{\A_{i}(x)}{\A_{i-1}(x)}\right) = \frac{1}{\hn} \cdot q_i(\B) \cdot \eta_i -  \frac{n}{\hn}\log \beta_i,
\end{align*}
where $\displaystyle{\eta_i = \frac{1}{2\hn} \cdot (m_i - q_i(\A_{i-1}))}$ and $\displaystyle{\beta_i = \frac{1}{\hn} \cdot \sum_{x \in \D} \exp(q_i(x) \eta_i)\A_{i-1}(x)}$.
Using the fact that $\exp(x) \le 1+x+x^2$ for $|x| \le 1$ and $|q_i(x)\eta_i| \le 1$,
\begin{align*}
    \beta_i \le & \frac{1}{\hn} \cdot \sum_{x \in \D}(1+q_i(x) \eta_i + q^2_i(x) \eta^2_i) \A_{i-1}(x) \\
    \le & \frac{1}{\hn} \cdot \sum_{x \in \D}(1+q_i(x) \eta_i + \eta^2_i) \A_{i-1}(x) \le 1 + \frac{1}{\hn} \cdot \eta_i q_i(\A_{i-1}) +  \eta^2_i.
\end{align*}
Then, we can rewrite $\Psi_i - \Psi_{i-1} =\frac{1}{\hn} \cdot q_i(\B) \cdot \eta_i - \log \beta_i$. Plugging the upper bound on $\beta_i$, we have
\begin{align*}
    \Psi_i - \Psi_{i-1} 
    \ge & \frac{\eta_i}{\hn} \cdot q_i(\B) - \left(\frac{\eta_i}{\hn} \cdot q_i(\A_{i-1}) +  \eta^2_i\right) 
    \ge \frac{1}{4\hn^2} \left((\ME_i - 4 \AE)^2 - \AE^2 \right),
\end{align*}
By rewriting the last inequality, we have
\begin{align*}
   \ME_i \le \sqrt{4\hn^2 \cdot (\Psi_i - \Psi_{i-1}) + \AE^2} + 4 \times \AE.
\end{align*}
Then, $\textrm{avg}_{i \le k} \ME_i \le \sqrt{4\hn^2 \cdot \textrm{avg}_{i \le k} (\Psi_i - \Psi_{i-1}) + \AE^2 } + 4 \AE \le 2\hn \cdot \sqrt{\frac{\log |\D|}{k}} \\+ 5 \AE$, which can be bounded by $O\left(\hn \cdot \sqrt{\frac{\log |\D|}{k}} + \frac{\log |\Q| \cdot \tDelta \sqrt{k \cdot \log(1/\delta)}}{\eps} \right)$.
By taking $k = \frac{\hn \cdot \epsilon \cdot \sqrt{\log |\D|}}{\tilde{\Delta} \cdot \log |\Q| \cdot \sqrt{\log(1/\delta)}}$, we can obtain the minimized error as $O(\sqrt{\hn \cdot \tDelta} \cdot \fup)$. Finally, recall that $\hn \leq n + O(\tDelta \cdot \lambda)$. This means that the error is at most $O((\sqrt{n \cdot \tDelta} + \tDelta \cdot \sqrt{\lambda}) \cdot \fup)$.

\section{Missing Proofs in Section~\ref{sec:join-as-one}}
\label{appendix:join-as-one}
\subsection{Upper Bound Proofs}
For neighboring instances $\I,\I'$,  $\big|\countsize(\I)- \countsize(\I')\big| \le \tilde{\Delta}$,
which follows from the definition of $\LS_\countsize(\I)$ and the non-negativity of $\TLap$.
\begin{proof}[Proof of Lemma~\ref{lem:two-table-DP}] 
It can be checked that $\LS_{\countsize}(\cdot)$ has global sensitivity of $1$.
Therefore, the DP guarantee of the truncated Laplace mechanism implies that $\tDelta$ (computed on Line 1) is $(\eps/2, \delta/2)$-DP. Applying the basic composition over this guarantee and the privacy guarantee in \Cref{thm:pmw}, we can conclude that the entire algorithm is $(\eps, \delta)$-DP.
\end{proof}

\begin{proof}[Proof of Lemma~\ref{lem:multi-table-DP}]
It follows from the definition that the global sensitivity of $\ln(RS^{\beta}_{\countsize}(\I))$ is at most $\beta$. Furthermore, observe that Line 2 can be rewritten as $\tDelta \gets \exp\left(\ln(RS^{\beta}_{\countsize}(\I)) + \TLap_{2\beta/\eps}^{\tau(\eps/2,\delta/2,\beta)}\right)$. Therefore, by the guarantee of truncated Laplace mechanism and the post-processing property of DP, we can conclude that $\tDelta$ is $(\eps/2, \delta/2)$-DP. Again, applying the basic composition of this guarantee and the privacy guarantee in \Cref{thm:pmw}, we can conclude that the entire algorithm is $(\eps, \delta)$-DP.
\end{proof}

\subsection{Lower Bound Proofs}

The main idea behind proving Theorem~\ref{the:lb-multi-table} is similar to that of Theorem~\ref{the:lb-two-table-2}, where one relation encodes the single table, and remaining tables ``amplify'' both sensitivity and join size by a $\Delta$ factor.

\begin{proof}[Proof of Theorem~\ref{the:lb-multi-table}]
Let $n= \frac{\OUT}{\Delta}$. From Theorem~\ref{the:lb-single-table}, there exists a set $\Q_\textsf{one}$ of queries on domain $\D$ for which any $(\epsilon, \delta)$-DP algorithm that takes as input a single-table instance $T \in \D$ and outputs an approximate answer to each query in $\Q_\textsf{one}$ within $\ell_\infty$-error $\alpha$ requires that $\alpha \ge \Tilde{\Omega}\left(\min\left\{n, \sqrt{n} \cdot \flo(\D,\Q_\textsf{one},\epsilon)\right\}\right)$.
For an arbitrary single-table instance $T: \D \to \mathbb{Z}^+$, we construct
a multi-table instance $\I$ for $\mathcal{H}$ of input size $m$, join size $\OUT$, and local sensitivity $\Delta$ as follows. We pick the relation with the smallest number of attributes to encode $T$. W.l.o.g., we assume that $|\x_1| = \min_{i} |\x_i|$. As $m \ge 2$, there must exist an attribute $y \in \x - \x_1$. Let $x \in \x_1$ be an arbitrary attribute. Let $k = \left|\x - \x_1\right|$.
\begin{itemize}[leftmargin=*]
    \item Set $\dom(x) = \D \times [n]$ for each $x \in \x_1$, and $\dom(y) = [\Delta^{1/k}]$ for each $y \in \x - \x_1$; 
    \item Let $R_1((a,b),\dots (a,b)) = \mathbf{1}[b \le T(a)]$ for all $a \in \D$ and $b\in [n]$.   
    \item For $i > 1$, let $R_i(t)=1$ for all $t \in \D_i$;
\end{itemize}
It can be easily checked that $\I$ has join size $\OUT$ and local sensitivity $\Delta$, and that two neighboring instances $T,T'$ result in neighboring instances $\I,\I'$.
Finally, let $\Q_1$ contain queries from $\Q_\textsf{one}$ applied on its first value of every attribute (i.e., $\Q_1 := \{q \circ \pi_{x,1} \mid q \in \Q_\textsf{one}\}$) for $x \in \x_1$, and let $\Q^i$ contain only a single query $q_\textsf{all-one}: \D_i \to \{+1\}$ for every $i \ge 2$.
The remaining argument follows exactly the same for the two-table case as in the proof of Theorem~\ref{the:lb-two-table-2}.
\end{proof}

\begin{theorem}
    For any $\Delta > 0$, $\epsilon > 0$, $\delta >0$, and $\gamma > 0$ such that $e^\epsilon \cdot \gamma + \delta < 1-\gamma$, there exists a family $\Q$ of linear queries such that any $(\epsilon, \delta)$-DP algorithm that takes as input an instance $\I$ with local sensitivity at most $\Delta$ and answers each query in $\Q$ to within error $\alpha$ with probability $1-\gamma$, must satisfy $\alpha \ge \Omega\left(\Delta\right)$.
\end{theorem}
\begin{proof}
    Let $\Q = \{\countsize\}$. Let $\I, \I'$ be neighboring instances of local sensitivity at most $\Delta$ such that $|\countsize(\I) - \countsize(\I')| \geq \Omega(\Delta)$. Suppose there is an $(\epsilon, \delta)$-DP algorithm $\cA$ that achieves error $\alpha < |\countsize(\I) - \countsize(\I')| / 2$ with probability at least $1 - \gamma$. Let $\tilde{\countsize}(\I), \tilde{\countsize}(\I')$ be the join sizes released for $\I, \I'$ respectively. Then, as $\cA$ preserves $(\epsilon, \delta)$-DP, it must be that
    \begin{align*}
        1 - \gamma &\leq \Pr\left(\tilde{\countsize}(\I) \in [\countsize(\I)-\alpha, \countsize(\I) + \alpha] \right) \\
        &\le e^\epsilon \cdot \Pr\left(\tilde{\countsize}(\I') \in [\countsize(\I)-\alpha, \countsize(\I) + \alpha]\right) + \delta \\
        &< e^\epsilon \cdot \left(1 - \Pr\left(\tilde{\countsize}(\I') \in [\countsize(\I')-\alpha, \countsize(\I') + \alpha]\right)\right) + \delta \\
        &\leq e^\eps \cdot \gamma + \delta < 1 - \gamma,
    \end{align*}
    a contradiction.
\end{proof}

\subsection{Worst-Case Error Bound}
We distinguish two cases below: (1) $R_i: \mathbb{D}_i \to \{0,1\}$; (2) $R_i: \mathbb{D}_i \to \mathbb{Z}^{\ge 0}$. For simplicity, we assume $\epsilon = \Theta(1)$ and $\delta = 1/n^c$ for some constant $c > 0$. Therefore, $\lambda = \Theta(1)$ and $\beta= \Theta(1)$.

In the first case, we recall the AGM bound~\cite{atserias2008size} on the maximum join size of a multi-way join. More specifically, let $\rho(\mathcal{H})$ be the fractional edge covering number of $\mathcal{H} $, which is the minimum value of $\sum_{i \in [m]} W_i$ subject to (1) $\sum_{i: x \in \x_i} W_i \ge 1$ for each $x \in \x$ and (2) $ W_i \in [0,1]$ for each $i \in [m]$. Then, $\countsize(\I) \le n^{\rho(\mathcal{H})}$ for any instance $\I$ of input size $n$. Now, we give an upper bound on the worst-case error. The residual sensitivity $RS^\beta_\countsize(\I)$ simply degenerates to~\cite{dong2021residual}: $\displaystyle{O\left(\max_{i \in [m]} \max_{E \subseteq [m] \smallsetminus \{i\}} T_{[m] \smallsetminus \{i\} \smallsetminus E}(\I)\right)= O\left(\max_{E \subsetneq [m]} T_E(\I)\right)}$.
So, $T_E(\I)$ is bounded by the maximum join size of $\mathcal{H}_{E,\partial E} = (\cup_{e \in E} e \smallsetminus \partial E, \{\x_i \smallsetminus \partial E: i \in E\})$. From the AGM bound, we have $T_E(\I) \le n^{\rho\left(\mathcal{H}_{E,\partial E}\right)}$. 
The worst-case error in Theorem~\ref{the:ub-multi-table} has a closed-form of $\displaystyle{O_{\lambda, \fup}\left(\sqrt{n^{\rho(\mathcal{H})} \cdot \max_{E \subsetneq [m]}n^{\rho\left(\mathcal{H}_{E,\partial E}\right)}}\right)}$. On the other hand, this is always smaller (at least not larger) than $O_{\lambda, \fup}(n^{\rho(\mathcal{H})})$, i.e., the maximum join size of the input join query.

In the second case, the AGM bound does not hold any more. A simpler tight bound on the maximum join size is $\Theta(n^m)$. Correspondingly, $\max_{E\subsetneq [m]}T_E(\I) \le n^{m-1}$. Putting everything together, the worst-case error in Theorem~\ref{the:ub-multi-table} has a closed-form of $\displaystyle{O_{\lambda, \fup}\left(n^{m-\frac{1}{2}}\right)}$.  
\section{Missing Proofs in Section~\ref{sec:uniformization}}
\label{appendix:uniformization}

\subsection{Two-Table Join}

\begin{lemma} \label{lem:partition-2table-dp}
\Cref{alg:partition-2table} is $(\eps, \delta)$-DP.
\end{lemma}

\begin{proof}
Define $\LS_\countsize(\I) = \max_{b\in \dom(B)}\{\deg^\I_{1,B}(b),\deg^\I_{2,B}(b)\}$. It is simple to observe that the global sensitivity of $\LS_\countsize$ is one. Furthermore, the output partition is simply a post-processing of the truncated Laplace mechanism, which satisfies $(\eps, \delta)$-DP.
\end{proof}

\begin{proof}[Proof of Lemma~\ref{lem:partition-2table-DP}]
From \Cref{lem:partition-2table-dp} the partition $\bbI$ is $(\eps/2,\delta/2)$-DP. Furthermore, {\sc TwoTable} (\Cref{alg:two-table}) is $(\eps/2,\delta/2)$-DP and is applied on disjoint parts of input data. Thus, by the parallel composition theorem and the basic composition theorem, we can conclude that the entire algorithm is $(\eps, \delta)$-DP.
\end{proof}

\begin{proof}[Proof of Theorem~\ref{the:up-2table-uniformize}]
Given an input instance $\I$ of the two-table join, let $\pi_1 = \left\{B^1_1, \dots, B_1^{\ell}\right\}$ be a fixed partition of $\dom(B)$, such that $b \in B^i_1$ if and only if $\max\{\deg_{1,B}(b), \deg_{2,B}(b)\} \in (\gamma_{i-1}, \gamma_i]$. Let $\pi_2 = \left\{B^1_2, \dots, B_2^{\ell}\right\}$ be the partition of $\dom(B)$ returned by Algorithm~\ref{alg:partition-2table}. If $b \in B^i_2$, 
$\max\{\deg_{1,B}(b), \deg_{2,B}(b)\} \in (\gamma_{i-1} - \lambda, \gamma_i]$. It is easy to see that $B^{i}_1\subseteq B^{i}_2 \cup B^{i+1}_2$. For simplicity, we denote the join size contributed by values in $B^{i}_1 \cap B^{i}_2$ and in $B^{i}_1 \smallsetminus B^{i}_2$ as $x_i, y_i$ respectively. Then, the cost of Algorithm~\ref{alg:partition-2table} under partition $\pi_2$ is $\sum_{i \in [\ell]}\sqrt{x_i + y_{i-1}} \cdot \sqrt{\lambda \cdot 2^{i}} \le \sum_{i \in [\ell]} (\sqrt{x_i} + \sqrt{y_{i-1}}) \cdot \sqrt{2^{i} \cdot \lambda} \le 2 \sum_{i \in [\ell]} \sqrt{x_i + y_i} \cdot \sqrt{2^{i} \cdot \lambda}$,
thus can be bounded by that under $\pi_1$.
\end{proof}

\begin{proof}[Proof of Theorem~\ref{the:lb-instance-optimality-2table}]
    Our proof consists of two steps: 
    \begin{itemize}[leftmargin=*]
        \item {\bf Step (1):} There exists a family $\Q^i$ of queries such that any $(\epsilon,\delta)$-DP algorithm that takes as input an instance of join size $\OUT^i$ and local sensitivity $\Delta_i = \Theta(2^i\cdot\lambda)$, and outputs an approximate answer to each query in $\Q^i$ to within error $\alpha^{i}$ must require $\displaystyle{\alpha_i \ge \tOmega\left(\min\left\{\OUT^i, \sqrt{\OUT^i \cdot 2^i \cdot \lambda} \cdot \flo\right\}\right)}$.
        \item {\bf Step (2):} There exists a family  $\Q$ of linear queries such that any $(\epsilon,\delta)$-DP algorithm that takes as input an instance $\I$ that conforms to $\overrightarrow{\OUT}$ and answers each query in $\Q$ to within error $\alpha$ must require
        $\displaystyle{\alpha \ge \tOmega\left(\max_i \min\left\{\OUT^i, \sqrt{\OUT^i \cdot 2^i \cdot \lambda } \cdot \flo\right\}\right)}.$
    \end{itemize}
    
    We first focus on {\bf step (1)} for an arbitrary $i \in [\lceil \log (\frac{n}{\lambda}) \rceil]$.
    Let $n_i$ be an arbitrary integer such that $n_i \cdot \lambda \cdot 2^{i-1} \le \OUT^i \le n_i \cdot \lambda \cdot 2^i$. Set $\Delta_i = \frac{\OUT^i}{n_i}$, where $\Delta_i \in (\lambda \cdot 2^{i-1}, \lambda \cdot 2^{i}]$. From Theorem~\ref{the:lb-single-table}, let $\Q^i_\textsf{one}$ be the set of hard queries on which any $(\epsilon,\delta)$-DP algorithm takes as input any single-table $T^i \in (\{0,1\}^d)^{n_i}$, and outputs an approximate answer to each query in $\Q^i_\textsf{one}$ to within error $\alpha$ must require $\alpha \ge \widetilde{\Omega}\left(\min\{n_i, \sqrt{n_i} \cdot \flo(\D,\Q^i_\textsf{one}, \epsilon)\}\right)$. For an arbitrary single-table $T^i \in (\{0,1\}^d)^{n_i}$, we can construct a two-table instance $(R^i_1, R^i_2)$ as follows:
    \begin{itemize}[leftmargin=*]
    \item Set $\dom(A) = \D$, $\dom(B) = \D \times [n]$ and $\dom(C) = [\Delta_i]$;
    \item Let $R^i_1(a,(b_1,b_2)) = \mathbf{1}[a = b_1 \cap b_2 \le T(a)]$ for all $a \in \dom(A)$ and $(b_1,b_2) \in \dom(B)$.
    \item Let $R^i_2 = 1$ for all $b \in \dom(B)$ and $c \in \dom(C)$. 
\end{itemize}
 
    It can be easily checked that $(R^i_1, R^i_2)$ has join size $\OUT^i$ and local sensitivity $\Delta_i$, and that two neighboring instances $T^i,T'^i$ result in neighboring instances $(R^i_1, R^i_2), ({R'}^{i}_1, {R'}^{i}_2)$ for the two-table join.
    Finally, let $\Q^i_1$ contain queries from $\Q^i_\textsf{one}$ applied on its first attribute (i.e., $\Q^i_1 := \{q \circ \pi_A \mid q \in \Q_\textsf{one}\}$), and let $\Q_2$ contain only a single query $q_\textsf{all-one}: \D_2 \to \{+1\}$.
 
    We use an argument similar to Theorem~\ref{the:lb-two-table-2}, showing that if there exists an $(\epsilon, \delta)$-DP algorithm that takes in a two-table instance of join size $\OUT^i$ and local sensitivity $\Delta_i$, and outputs an approximate answer to each query in $\Q^i_\textsf{two}$ to within error $\alpha$, there exists an $(\epsilon, \delta)$-DP algorithm that takes an arbitrary single-table $T_i \in (\{0,1\})^{n_i}$, and outputs an approximate answer to each query in $\Q^i_\textsf{one}$ to within error $\frac{\alpha^i}{\Delta^i} \ge \tOmega\left(\min\{n_i, \sqrt{n_i} \cdot \flo(\D,\Q^i_\textsf{one}, \epsilon)\}\right)$; hence $\alpha^i \ge \tOmega\left(\min\left\{\OUT^i, \sqrt{\OUT^i \cdot 2^i \cdot \lambda} \cdot \flo\right\}\right)$.
    
    \paragraph{Step (2).} From Theorem~\ref{the:lb-single-table}, let $\Q_\textsf{two}$ be the family of linear queries over $\D_1 \times \D_2$, such that $\displaystyle{\Q_\textsf{two} = \left\{\cup_{i \in [m]} q_i: q_i \in \Q^{i}_\textsf{two}, \forall i \in [\ell]\right\}}$. Consider an  $(\epsilon, \delta)$-DP algorithm $\mathcal{A}$ takes as input an instance that conforms to $\overrightarrow{\OUT}$ and answers each query in $\Q_\textsf{two}$ to within error $\alpha$. If $\displaystyle{\alpha \le \tO\left(\max_i \min\left\{\OUT^i, \sqrt{\OUT^i \cdot 2^i \cdot \lambda} \cdot \flo\right\}\right)}$, there exists an $(\epsilon, \delta)$-DP algorithm that takes an input an instance of join size $\OUT^i$ and local sensitivity $\Delta_i$, and outputs an approximate answer to each query in $\Q^i_\textsf{two}$ for some $i$, contradicting {\bf Step (1)}.
 \end{proof}

 \subsection{Hierarchical Join}

   \begin{proof}[Proof of Lemma~\ref{lem:hierarchical-degree-function}]
   Let $r$ be the root of $\mathcal{T}$ and let $\textsf{path}(r,x)$ be the set of attributes on the path from $r$ to $x$.  The proof has three steps.
   
   {\bf Step 1.}  We show that for any $T_{E,\y}$ involved, if $T_{E, \y}$ falls into case (2.1), then $\partial E \subseteq \y$, and if $T_{E,\y}$ falls into case (2.2), then $\partial E = \y$. Initially, either case holds for $T_{E,\partial E}$. Moreover, we observe that the cases (2.1) and (2.2) are invoked exchangeably, i.e., $T_{E', \y \cap ({\bigvee_{E'}})}$ will fall into case (2.2), and $T_{E,\bigwedge_E}$ will fall into case (2.1). Next, we show that this property is preserved in each recursive step: 
    \begin{itemize}[leftmargin=*]
        \item $\mathcal{H}_{E,\y}$ is disconnected. By hypothesis, assume $\partial E \subseteq \y$. Consider an arbitrary connected subquery $E'' \in \mathcal{C}_E$. For any pair of $i \in E''$ and $j \notin E$, $\x_i \cap \x_j \in \y$. For any pair of $i \in E''$ and $j \in E \smallsetminus E''$, $\x_i \cap \x_j \in \y$. This way, $\partial E'' \subseteq \y$. Together with $\partial E'' \subseteq \bigwedge_{E''}$, we obtain $\partial E'' \subseteq \y \cap (\bigwedge_{E''})$. On the other hand, for each attribute $x \in \y \cap (\bigwedge_{E''})$, there must exist some $i \in E''$ with $x \in \x_i$ and some $j \notin E$ such that $x \in \x_j$. This way, $\y \cap (\bigwedge_{E''}) \subseteq \partial E''$. Hence, $\partial E'' = \y \cap (\bigwedge_{E''})$ for $T_{E'', \y \cap (\bigwedge_{E''})}$.
        \item $\mathcal{H}_{E,\y}$ is connected. By hypothesis, assume $\partial E = \y$. As $\y \subsetneq \bigwedge_E$, $\partial E \subseteq \bigwedge_E$. Hence, $T_{E,\bigwedge_E}$ falls into case (2.1) and has $\partial E \subseteq \bigwedge_E$.  
    \end{itemize}
    Note that $\degg_{E}(\y)$ is only introduced in case (2.2), hence $\partial E = \y$. 
    
    {\bf Step 2.} From Definition~\ref{def:degree-function}, $\y \subseteq \bigwedge_E$. In Case (1) with $|E|=1$, we must have $\y \neq \x_i$, thus $\y \subsetneq \x_i$. In Case (2.2), $\y \subsetneq \bigwedge_E$ follows the fact that $\mathcal{H}_{E,\y}$ is disconnected. Together, $\y \subsetneq \bigwedge_E$.

    {\bf Step 3.} We first note that $E \subseteq \textsf{atom}(x)$. Suppose $i \in \textsf{atom}(x) \smallsetminus E$. We have $\textsf{path}(r,x) \in \partial E$, hence $\textsf{path}(r,x) \in \y$ from {\bf Step 1}. This way, $\y = \bigwedge_E$ contradicts {\bf Step 2}. Hence, $\textsf{atom}(x) \subseteq E$. Together, we have $\textsf{atom}(x) = E$. Moreover, as $\y \subseteq \bigwedge_E$, there must be $\y \subseteq \textsf{path}(r,x')$. Suppose $x' \notin \y$. Then, there exists some relation $i \notin E$ with $\textsf{path}(r,x') \in \x_i$. In this way, $x' \in \partial E$ and therefore $x' \in \y$, yielding a contradiction of $x' \notin \y$. Hence, $\y =\textsf{path}(r,x')$. 
\end{proof}

\begin{proof}[Proof of Lemma~\ref{lem:partition-hierarchical}]
     We prove the first property by induction. In the base case with $|\mathbb{I}|=1$, these properties hold. 
     Consider an arbitrary iteration of {\bf while} loop in Algorithm~\ref{alg:uniformize-hierarchical}. By hypothesis, all sub-instances in $\mathbb{I}$ have disjoint join results, and their union is $\textsf{Join}^{\mathbb{I}}$. Then, it suffices to show that {\sc Decompose}$(\I, x)$ generates a set of sub-instances of $\I$, 
     such that they have disjoint join results, and their union is $\textsf{Join}^{\I}$. The disjointness is easy to show since $(R_{j,i})_{i \in [\ell]}$ forms a partition of $R_j$, for every $j \in \textsf{atom}(x)$. The completeness follows the partition of $\dom(\y)$ and definition of $R_{j,i}$. 
     
     For the second property, we show that each tuple $t \in R_i$ appears in $O(\ell^c)$ sub-instances, where $c = \sum_{x \in \x} |\textsf{atom}(x)|$. In an invocation of {\sc Partition-Hierarchical}$(\I,x)$, $t$ appears in $O\left(\ell \right)$ sub-instances if $i \notin \textsf{atom}(x)$, and only one sub-instance otherwise. Overall, the procedure {\sc Decompose} will be invoked on every non-leaf node in $\mathcal{T}$, thus
     tuple $t \in R_i$ appears in 
     $O\left(\prod_{x \in \x} \ell \right) =O\left( \ell^{|\x|} \right)$ sub-instances.
     
     Finally, we show that each sub-instance corresponds to a degree characterization.  Let us focus on an arbitrary relation $\x_j$ in Algorithm~\ref{alg:uniformize-hierarchical}. For simplicity, let $\langle x_1, \dots,x_k\rangle$ be the root-to-node path corresponding to $\x_j$. Also, let $\y_1,\dots, \y_k$ be the set of ancestors of $x_1, \dots, x_k$ respectively. When {\sc Decompose}$(\I,x_1)$ is invoked, the sub-instance $\I'$ with 
     $\deg^{\I'}_{\textsf{atom}(x_1), \y_1}(t) \in \left(\lambda \cdot 2^{i-1}, \lambda \cdot 2^{i}\right]$
     for any $t \in \dom(\y_1)$, corresponds to $\sigma(\textsf{atom}(x_1), \y_1)=i$. In the subsequent invocations of {\sc Decompose}, tuples with the same value $t \in \dom(\y_1)$ always fall into the same sub-instance, hence this property holds. 
    Thus, each sub-instance returned by Algorithm~\ref{alg:uniformize-hierarchical} corresponds to a distinct degree characterization. 
\end{proof}

\begin{proof}[Proof of Lemma~\ref{lem:partition-hierarchical-DP}]
    Consider two neighboring instances $\I$ and $\I'$. Note that $|\deg^\I_{E,\y}(t) - \deg^{\I'}_{E,\y}(t)| \le 1$ holds for any $E \subseteq [m]$, $\y \subseteq \bigwedge_E$, and $t \in \dom(\y)$. 
    Each tuple in $R_i$ contributes to at most $|\x_i|$ degrees, i.e., $\deg_{\textsf{atom}(x_i), \y}(\cdot)$. Hence, $\deg^\I\approx_{(c'\cdot \epsilon, c'\cdot\delta)} \deg^{\I'}$, where $c' = \max_{i\in[m]} |\x_i|$ is a join-query-dependent quantity.
    
    Moreover, for each sub-instance $\I^\sigma$ returned, $\F^{\sigma} \approx_{(3\epsilon, 3\delta)} \F'^{\sigma}$ implied by Lemma~\ref{lem:two-table-DP}.
    As each tuple participates in at most $O(\ell^c)$ sub-instances,  $\bigcup_{\sigma} \F^{\sigma} \approx_{(3\ell^c\cdot \epsilon, 3 \ell^c \cdot \delta)} \bigcup_{\sigma} \F'^{\sigma}$ implied by the group privacy. Putting everything together and using basic composition, we conclude that Algorithm~\ref{alg:uniformization-framework} is $(O(\ell^c \cdot \epsilon), O(\ell^c \cdot \delta))$-DP.
\end{proof} 

Given a degree configuration $\sigma$, we can upper bound $T_E(\I)$ as $T^\sigma_E$ and $\RS_\countsize(\I)$ as $\RS^\sigma_\countsize$, using Definition~\ref{def:residual}. Similar to the two-table case, we define the {\em uniform partition} of an instance $\I$ using true degrees as $\pi^* = \left\{\I^\sigma_{\pi^*}: \sigma \textrm{ is a degree configuration}\right\}$. Similarly, the error achieved by the partition based on noisy degrees can be bounded by the uniform one:
\begin{theorem}
\label{the:up-hierarchical-uniformize}
    For any hierarchical join $\H$ and an instance $\I$, a family $\Q$ of linear queries, and $\epsilon > 0$, $\delta > 0$, there exists an algorithm that is $(\epsilon, \delta)$-DP, and with probability at least $1- 1/\textsf{poly}(|\Q|)$ produces $\F$ such that all queries in $\Q$ can be answered to within error:
    \[\alpha = O\left( \left(\sum_{\sigma}  \sqrt{\countsize\left(\I^{\sigma}_{\pi^*}\right) \cdot RS^{\sigma}_\countsize \cdot \lambda} + RS^{\sigma}_\countsize \cdot \lambda\right) \cdot \fup \right),  \]
    where $\sigma$ is over all degree configurations of $\mathcal{H}$, $\I^\sigma_{\pi^*}$ is the sub-instance characterized by $\sigma$ under the uniform partition $\pi^*$, and $\RS^\sigma_\countsize$ be the residual sensitivity derived for instance characterized by $\sigma$.
\end{theorem}

Let $\overrightarrow{\OUT} = \left\{\OUT^\sigma \in \mathbb{Z}^\ge 0: \sigma \textrm{ is a degree configuration}\right\}$ be the join size distribution over the uniform partition $\pi^*$. 
An instance $\I$ conforms to $\overrightarrow{\OUT}$ if $\countsize(\I^\sigma_{\pi^*}) = \Theta(\OUT^\sigma)$, for every degree configuration $\sigma$, where $\I^\sigma_{\pi^*}$ is the sub-instance of $\I$ under $\sigma, \pi^*$. Extending the lower bound argument of Theorem~\ref{the:lb-instance-optimality-2table}, we obtain the following parameterized lower bound for hierarchical queries.

\ifbeforepasinchange
\else


\begin{theorem}
\label{the:lb-hierarhical-uniformize}
    Given a hierarchical join $\H$ and an arbitrary parameter $\overrightarrow{\OUT}$,
    for every sufficiently small $\eps >0$, $n_D \geq \OUT^{O(1)}$ and $n_Q \geq (\OUT \cdot \log n_D)^{O(1)}$, there exists a family $\Q$ of queries of size $n_Q$ on domain $\D$ of size $n_D$ such that any $(\epsilon, 1/n^{\omega(1)})$-DP algorithm that takes as input a multi-table instance over $\H$ of input size at most $n$ while conforming to $\overrightarrow{\OUT}$, and outputs an approximate answer to each query in $\Q$ to within error $\alpha$, must satisfy 
    \[\alpha \ge \tOmega\left(\max_{\sigma} \min\left\{\OUT^{\sigma}, \sqrt{\OUT^{\sigma}}\cdot \sqrt{\LS^\sigma_\countsize} \cdot \flo\right\}\right), \]
    where the maximum is over all degree configurations $\sigma$ of $\mathcal{H}$ and $\displaystyle{\LS^\sigma_\countsize = \max_{i \in [m]} T^\sigma_{[m] \smallsetminus \{i\}}}$ is the local sensitivity of $\countsize(\cdot)$ under $\sigma$.
\end{theorem}
\fi